\documentclass[journal]{IEEEtran}

\usepackage[pdftex]{graphicx}
\usepackage{amsmath}
\usepackage{array}
\usepackage{url}
\usepackage{multirow,makecell}
\usepackage{color,soul}

\begin{document}

\title{Degraded Reference Image Quality Assessment}

\author{Shahrukh~Athar,~\IEEEmembership{Member,~IEEE,}
        and~Zhou~Wang,~\IEEEmembership{Fellow,~IEEE}
\thanks{The authors are with the Department of Electrical and Computer Engineering, University of Waterloo, Waterloo, ON N2L 3G1, Canada (e-mail: \{shahrukh.athar, zhou.wang\}@uwaterloo.ca).}}%

\maketitle

\begin{abstract}
In practical media distribution systems, visual content usually undergoes multiple stages of quality degradation along the delivery chain, but the pristine source content is rarely available at most quality monitoring points along the chain to serve as a reference for quality assessment. As a result, full-reference (FR) and reduced-reference (RR) image quality assessment (IQA) methods are generally infeasible. Although no-reference (NR) methods are readily applicable, their performance is often not reliable. On the other hand, intermediate references of degraded quality are often available, e.g., at the input of video transcoders, but how to make the best use of them in proper ways has not been deeply investigated. Here we make one of the first attempts to establish a new paradigm named degraded-reference IQA (DR IQA). Specifically, we lay out the architectures of DR IQA and introduce a 6-bit code to denote the choices of configurations. We construct the first large-scale databases dedicated to DR IQA and will make them publicly available. We make novel observations on distortion behavior in multi-stage distortion pipelines by comprehensively analyzing five multiple distortion combinations. Based on these observations, we develop novel DR IQA models and make extensive comparisons with a series of baseline models derived from top-performing FR and NR models. The results suggest that DR IQA may offer significant performance improvement in multiple distortion environments, thereby establishing DR IQA as a valid IQA paradigm that is worth further exploration. 
\end{abstract}


\section{Introduction}
\label{sec:Intro}

\begin{figure}[t!]
	\centering
	\includegraphics[width=0.85\linewidth]{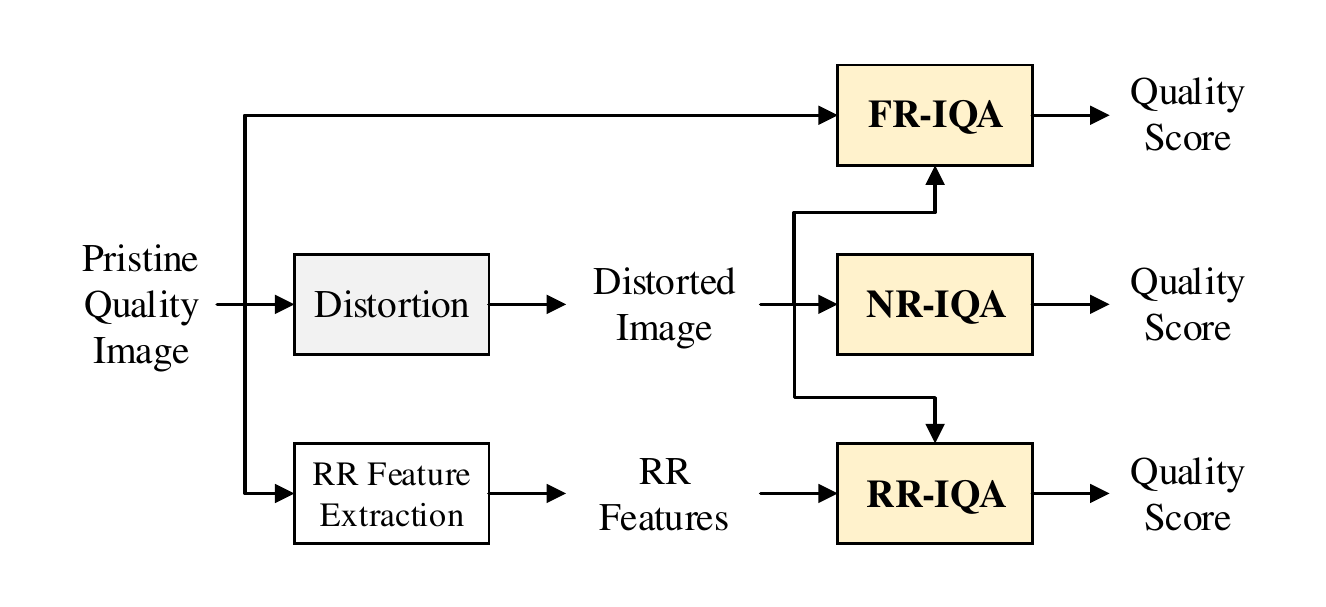}
	\vspace{-2mm}
	\caption{General framework of FR, RR and NR IQA.}
	\label{FRNQA}
	\vspace{-2mm}
\end{figure}

\begin{figure}[h!]
	\centering
	\vspace{-4mm}
	\includegraphics[width=1.0\linewidth]{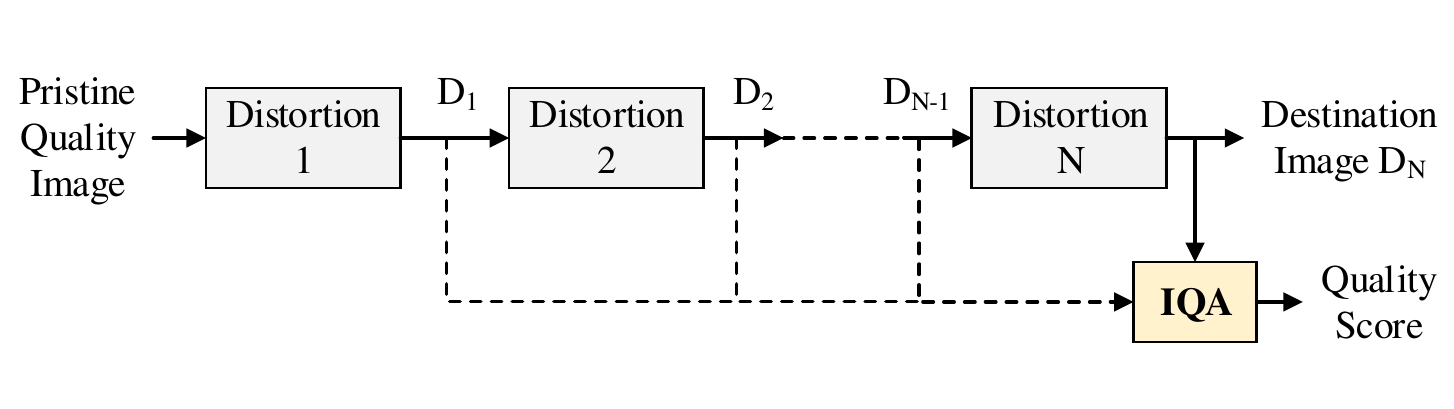}
	\vspace{-10mm}
	\caption{The framework of practical media distribution systems.}
	\label{MDQA}
\end{figure}

\IEEEPARstart{T}{he} goal of objective Image Quality Assessment (IQA) methods is to predict the quality of images as perceived by human eyes. Based on the accessibility to pristine reference content, they are traditionally classified into three paradigms, namely \textit{full-reference} (FR), \textit{reduced-reference} (RR) and \textit{no-reference} (NR) or \textit{blind} IQA (BIQA) \cite{iqa_book,rrnrSPM}, as illustrated in Fig. \ref{FRNQA}. In the literature, FR, RR, and NR IQA algorithms are usually tested and at times trained on image databases where each distorted image has undergone a single (often simulated) stage of distortion. This is in clear contrast to real-world visual content distribution scenarios, as illustrated in Fig. \ref{MDQA}, where visual content undergoes multiple stages of distortions before reaching its destination. For example, most consumer cameras and camcorders, including mobile phone cameras, store captured content using lossy compression standards such as JPEG and H.264. When these images and videos are uploaded to a social networking or video-sharing website, they usually undergo further rounds of lossy transcoding~\cite{youtube,facebook} for onward delivery to viewers. This means two stages of compression. For another example, an image or video maybe contaminated by noise or blur during acquisition. The camera will store this content in compressed form which may be followed by further compression during its distribution. This essentially means blur or noise contamination followed by compression. Compressed medical images provide another example of content afflicted by multiple distortion stages. It is known that magnetic resonance (MR), computed tomography (CT), and ultrasound images are affected by different types of noise \cite{MRnoise,CTnoise,USnoise}. With the rapid increase in the resolution and volume of medical images and with the emergence of tele-medicine, it is now desirable to largely reduce the data rate by lossy image compression as long as it does not affect the diagnostic quality \cite{medcompress1,medcompress2}. This leads to a distortion combination of noise followed by compression. Compressed astronomical images provide yet another example of noise followed by compression since such images are contaminated by noise \cite{astroimages}. Thus, even if we start with a pristine reference image, it may be affected by multiple stages of distortions by the time it reaches the end user. The requirement for IQA methods capable of handling multiple simultaneous distortions is not new~\cite{md_1960sRef}, but remains a major challenge~\cite{md_sevench}.

In practical media delivery systems, access to pristine reference images in the middle of the delivery chain is either extremely rare or altogether nonexistent. This, coupled with the multiple distortion nature of such systems, makes the use of FR and RR IQA infeasible. While NR IQA methods are readily applicable, most NR methods are trained and tested on databases that have images with a single stage of distortion, and their performance lags behind that of FR IQA methods \cite{eval_Waterloo}. Efforts have been made recently to design NR IQA methods that handle multiply distorted images. SISBLIM \cite{md_sisblim_db_mdid2013} is a training-free metric designed for singly and multiply distorted images through the fusion of estimates of noise, blur, JPEG compression, and joint effects. BoWSF \cite{md_BoWSF} selects features sensitive to different distortion types, which are encoded through a Bag-of-Words model and mapped to a quality score. LQAF \cite{md_lqaf} uses Support Vector Regression (SVR) to map features such as phase congruency, gradient magnitude, gray level gradient co-occurrence matrix and the contrast sensitivity function to quality scores. An enhanced and multi-scale version of LQAF, called MS-LQAF is proposed in \cite{md_mslqaf}. The training-based GWHGLBP \cite{md_gwhglbp} uses the gradient-weighted histogram of the local binary pattern (LBP) generated on the gradient map of the distorted image to capture the effects of multiple distortions. Jet-LBP \cite{md_jetLBP} uses color Gaussian jets to generate feature maps from a distorted image. The LBP is applied to these feature maps, followed by a weighted histogram that is mapped to quality scores through SVR. MUSIQUE \cite{md_musique} performs distortion identification followed by distortion parameter estimation and score generation. Nevertheless, we showed in \cite{md_waterloo17} that NR IQA methods generally perform unsatisfactorily when dealing with multiply distorted images, especially when the distortion types vary and with high distortion levels at earlier stages.

In addition to the performance issues, another major limitation of NR IQA methods is that they are incapable of incorporating the mid-stage distorted images along the media delivery chain, shown as $D_1$ to $D_{N-1}$ in Fig.~\ref{MDQA}. For example, at the input of a video transcoder, typically a compressed video stream of degraded quality, is often available but not used by NR IQA methods to assess the video stream at transcoder output. The question is: How to best utilize the auxiliary information of the mid-stage images of degraded quality to assess the quality of the final multiply distorted images? We term this problem \textit{degraded-reference} IQA (DR IQA).

This work is by no means the first attempt to tackle the DR IQA problem. A pioneering work is the \textit{corrupted-reference} (CR) IQA scheme laid out in the context of an image restoration problem~\cite{cr_icip12awgn,cr_icip14poisson,cr_tip19}, where the quality of a denoised image with respect to an absent pristine reference image is estimated by using a Gaussian or Poisson noise contaminated corrupted reference image. A recent interesting work targeting at DR IQA in image restoration scenarios learns a reference space for DR images by knowledge distillation from pristine images~\cite{dr_iccv21}. These are instantiations of Type-010010 DR IQA based on the categorization we will introduce in Section~\ref{sec:DRIQAparadigm}. In our earlier work \cite{md_waterloo17}, we show that Type-100100 DR IQA offers the potential to substantially elevate the performance of quality prediction against two baselines: FR IQA between the degraded-reference and final distorted images, and NR IQA of the final distorted image. The two-step quality assessment (2stepQA) scheme \cite{md_2stepQA_conf,md_2stepQA_jrnl} represents a series of Type-001100 DR IQA instantiations, where many combinations of FR methods (PSNR, MSSSIM \cite{fr_msssim}, FSIM \cite{fr_fsim}, VSI \cite{fr_vsi}) between the degraded-reference and final distorted images, and NR methods (NIQE \cite{nr_niqe}, BRISQUE \cite{nr_brisque}, CORNIA \cite{nr_cornia}, PQR \cite{nr_cnn_pqr}) on the degraded reference images, have demonstrated great promises, though it does not take into account how different distortions behave in conjunction with each other. Other types of DR IQA architectures (as elaborated in Section~\ref{sec:DRIQAparadigm}), to the best of our knowledge, have not been attempted in the literature. 

In this paper, we make one of the first attempts to establish DR IQA as a new IQA paradigm. We lay out and discuss 53 potential architectures for DR IQA (Section~\ref{sec:DRIQAparadigm}). We construct two new large-scale synthetically annotated datasets dedicated to DR IQA (Section \ref{sec:DRIQAdatabases}). We study the behaviors in multiple simultaneous distortions and make some interesting observations not reported before (Section \ref{sec:MDbehaveAnalysis}). Based on these observations, we develop novel DR IQA models (Section \ref{sec:DRIQAmodeling}) and make extensive comparisons with baseline models derived from top-performing FR and NR models (Section \ref{sec:PerfEval}). Finally, we conclude that the DR IQA paradigm offers great potentials and is worth further exploration in the future (Section \ref{sec:Concl}).

\section{Degraded Reference IQA Architectures}
\label{sec:DRIQAparadigm}

\begin{figure}[t!]
	\centering
	\includegraphics[width=1.0\linewidth]{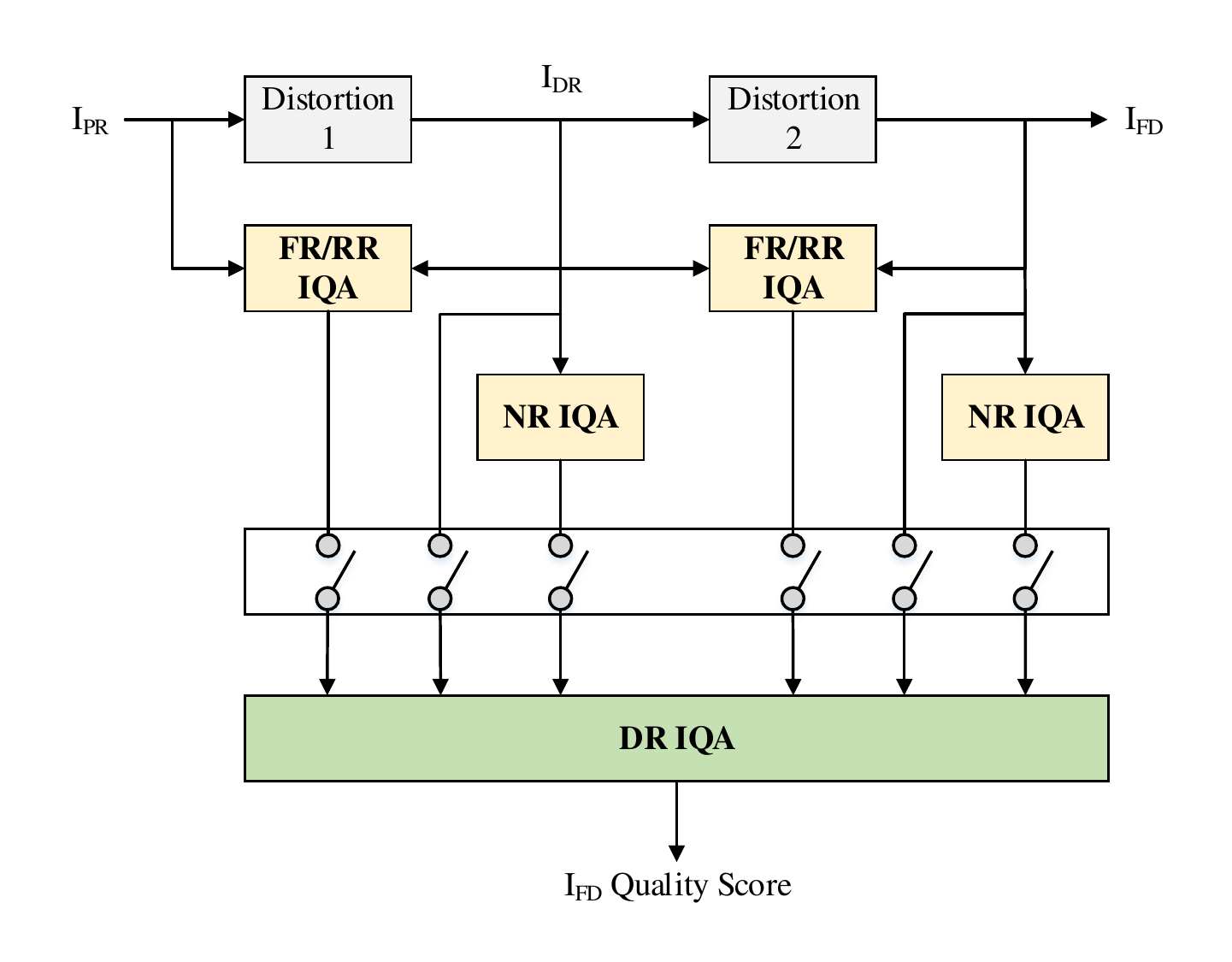}
	\vspace{-6mm}
	\caption{General architectures of DR IQA. Configurations determined by the statuses of six switches.}
	\label{fig:DRIQA_Architectures}
	\vspace{-3mm}
\end{figure}

Although in practice images and videos may undergo many stages of distortions, a logical starting point to study DR IQA is to focus on the case of two stages of distortions. This allows for a thorough discussion about all potential ways to perform DR IQA without missing the main issues that may be encountered in the cases of many stages of distortions. In a two-stage distortion pipeline, a source of a \textit{pristine reference} (PR) image undergoes stage-1 distortion, leading to a \textit{degraded reference} (DR) image, which subsequently undergoes stage-2 distortion and results in a \textit{final distorted} (FD) image, as shown at the top part of Fig.~\ref{fig:DRIQA_Architectures}. In DR IQA, both DR and FD images are assumed available, while the PR image is generally not accessible. However, in certain circumstances, information regarding the degradation from the PR to DR images may be made available to the DR IQA module, which often helps improve the performance the DR IQA algorithms.

Although traditional FR/RR/NR IQA methods alone do not directly provide adequate solutions to the DR IQA problem, they may be employed as key components in the total solution. Depending on how they are utilized, there may be many configurations of DR IQA architectures, which are summarized in Fig.~\ref{fig:DRIQA_Architectures}. In particular, the FR/RR IQA computed between the PR and DR images, and between the DR and FD images, the NR IQA computed from the DR and FD images, together with the DR and FD images themselves, may all be part of the input to the DR IQA module, which produces a single quality score of the FD image. The inputs to the DR IQA module may be controlled by 6 switches, each of which may be turned on or off independently, resulting in 64 potential configurations of DR IQA architectures, each represented by a 6-bit architecture code, where a bit of 1 or 0 denotes a switch being ON or OFF, respectively, and the order of the switches are given in Fig.~\ref{fig:DRIQA_Architectures}.


\begin{table}[t!]
	\scriptsize
	\caption{DR IQA architecture type groups and switch code}
	\vspace{-2mm}
	\centering
	\begin{tabular}{ c | c }
		\hline
		Architecture & \multirow{2}{*}{Architecture Switch Code} \\
		Type & \\ \hline
		\multirowcell{4}{Type-0 \\ Group} & 000100, 000101, 000110, 000111, 001001, 001010, 001011, \\
		& 001100, 001101, 001110, 001111, 010001, 010010, 010011, \\
		& 010100, 010101, 010110, 010111, 011001, 011010, 011011, \\
		& 011100, 011101, 011110, 011111 \\ \hline
		\multirowcell{4}{Type-1 \\ Group} & 100001, 100010, 100011, 100100, 100101, 100110, 100111, \\
		& 101001, 101010, 101011, 101100, 101101, 101110, 101111, \\
		& 110001, 110010, 110011, 110100, 110101, 110110, 110111, \\
		& 111001, 111010, 111011, 111100, 111101, 111110, 111111 \\ \hline
		\multirowcell{2}{Invalid \\ Group} & 000000, 000001, 000010, 000011, 001000, 010000, 011000, \\
		& 100000, 101000, 110000, 111000 \\ \hline
	\end{tabular}
	\label{Table_DRIQA_Arch}
	\vspace{-2mm}
\end{table}

The 64 potential architectures may be further classified into three groups. In the case that the PR image is completely inaccessible (even in gauging the FR/RR quality degradation from the PR image to the DR image), the first bit of the architecture code equals 0. Therefore, we categorize these architectures into the Type-0 group. Correspondingly, when information regarding the PR image is accessed through the FR/RR measure between PR and DR images, we classify the architectures into the Type-1 group. There are also invalid configurations for DR IQA when either the DR or the FD image is completely inaccessible to the DR IQA module. The classifications and the corresponding architecture codes are listed in Table~\ref{Table_DRIQA_Arch}, where there are 25, 28, and 11 architectures in the Type-0, Type-1 and invalid groups, respectively.

In practice, not all of the 53 (25+28) valid DR IQA architectures are equally favored, and the choice is likely dependent on the application scenarios. For instance, FR IQA methods are generally more reliable than other IQA approaches, and thus Type-1***** or Type-***1** are desirable. But in practical video delivery pipelines, obtaining both PR and DR videos or both DR and FD videos at the same monitoring points may not be easy, and even when the condition is met, the videos are often not aligned along the temporal dimension. For another example, when no existing FR/RR/NR IQA method is available, then Type-010010 is the only option, but this would require a completely new design of the DR IQA approach. More sophisticated cases may also occur with certain mixtures of distortion types in the two stages of distortions, which will be elaborated in more details in Section~\ref{sec:MDbehaveAnalysis}.

\begin{figure}[t!]
	\centering
	\includegraphics[width=0.85\linewidth]{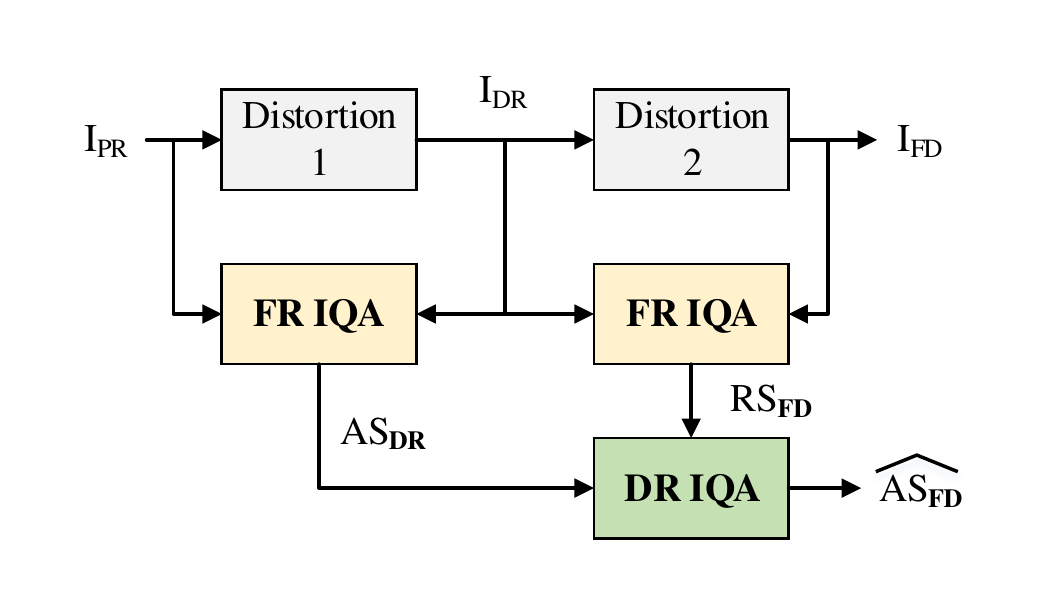}
	\vspace{-6mm}
	\caption{Scenario 1: Type-100100 DR IQA architecture.}
	\label{DRIQA_Scenario1}
	\vspace{-4mm}
\end{figure}

In the rest of this paper, we focus on two scenarios corresponding to two DR IQA architectures, based on which we study the behaviors of two-stage quality variations under different distortion combinations in Section~\ref{sec:MDbehaveAnalysis} and develop DR IQA models in Section~\ref{sec:DRIQAmodeling}. The first scenario corresponds to Type-100100 DR IQA, and this architecture is given in Fig.~\ref{DRIQA_Scenario1}. Since FR methods provide the most reliable quality prediction performance \cite{eval_Waterloo} and the PR image is assumed to have perfect quality, to simplify the analysis, we define the \textit{absolute scores} (AS) of a test image as the FR measure between the test image and its corresponding PR image. In the case of two distortion stages, we may compute the AS scores of the DR and FD images as Eq. \ref{AS_DR} and \ref{AS_FD}:
\begin{equation}
	\centering
	{\text{AS}_{\text{DR}} = \text{FR}(\text{I}_{\text{PR}}, \text{I}_{\text{DR}}),
		\label{AS_DR}}
\end{equation}
\begin{equation}
	\centering
	{\text{AS}_{\text{FD}} = \text{FR}(\text{I}_{\text{PR}}, \text{I}_{\text{FD}}),
		\label{AS_FD}}
\end{equation}
respectively, where $\text{I}_{\text{PR}}$, $\text{I}_{\text{DR}}$, and $\text{I}_{\text{FD}}$ are the PR, DR and FD images, respectively. The third possible FR comparison is between the DR and FD images, which assesses the quality of the FD image relative to the DR image. Thus, we regard it as the \textit{relative score} (RS) of the FD image:
\begin{equation}
	\centering
	{\text{RS}_{\text{FD}} = \text{FR}(\text{I}_{\text{DR}}, \text{I}_{\text{FD}}).
		\label{RS_FD}}
\end{equation}
Ideally, if FR IQA is fully trusted, then $\text{AS}_{\text{FD}}$ yields the best quality estimate of the FD image, but cannot be computed as the PR image is not available, at the end user level. Therefore, in Scenario 1, which assumes the availability of the PR image early on in the media distribution system, the goal of the DR IQA module in Fig.~\ref{DRIQA_Scenario1} is to make the best prediction of $\text{AS}_{\text{FD}}$ using $\text{AS}_{\text{DR}}$ and $\text{RS}_{\text{FD}}$, i.e.,
\begin{equation}
	\centering
	{\widehat{\text{AS}_{\text{FD}}} = f(\text{AS}_{\text{DR}},\text{RS}_{\text{FD}}),
		\label{DRIQA_Sc1}}
\end{equation}
where $f$ is the prediction function of the DR IQA module. Scenario 1 is practically applicable only when $\text{AS}_{\text{DR}}$ is pre-computed at the first distortion stage and transmitted as side information with the DR image to the second distortion stage. Apparently, transferring $\text{AS}_{\text{DR}}$ would require minor protocol changes of the media delivery system.

\begin{figure}[t!]
	\centering
	\includegraphics[width=0.85\linewidth]{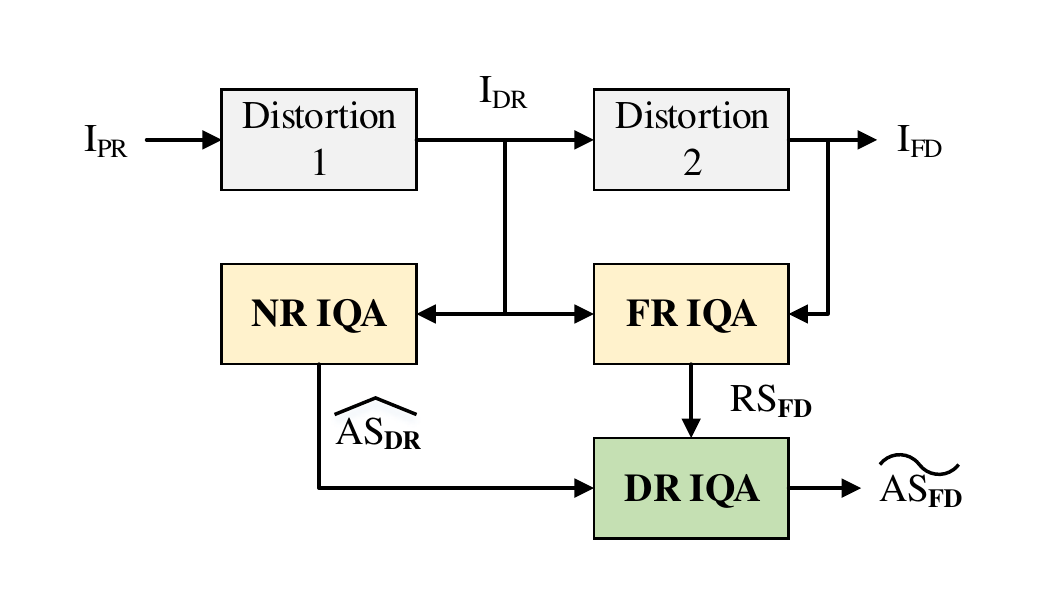}
	\vspace{-6mm}
	\caption{Scenario 2: Type-001100 DR IQA architecture.}
	\label{DRIQA_Scenario2}
	\vspace{-4mm}
\end{figure}

The second scenario follows Type-001100 DR IQA architecture, as shown in Fig.~\ref{DRIQA_Scenario2}, where the PR image is completely inaccessible, and thus $\text{AS}_{\text{DR}}$ cannot be computed. Instead, an NR IQA method is employed to produce an estimate 
\begin{equation}
	\centering
	{\widehat{\text{AS}_{\text{DR}}} = \text{NR}(\text{I}_{\text{DR}}).
		\label{AS_DRest}}
\end{equation}
Correspondingly, the DR IQA module is designed to predict $\text{AS}_{\text{FD}}$ by using NR IQA predicted $\widehat{\text{AS}_{\text{DR}}}$ and $\text{RS}_{\text{FD}}$:
\begin{equation}
	\centering
	{\widetilde{\text{AS}_{\text{FD}}} = g(\widehat{\text{AS}_{\text{DR}}},\text{RS}_{\text{FD}}),
		\label{DRIQA_Sc2}}
\end{equation}
where $g$ is the prediction function of the DR IQA module and may take a similar form as $f$ in Eq.~\ref{DRIQA_Sc1}. Unlike the first scenario, there is no need here to make any change to the existing media distribution system, and the DR IQA systems can be deployed as passive quality monitoring probes. This makes Scenario 2 both practically applicable and readily deploy-able.

\section{Multiple Distortions: Database Construction and Behavior Analysis}

\subsection{Database Construction}
\label{sec:DRIQAdatabases}

Most existing IQA datasets of multiply distorted content such as LIVE MD \cite{db_livemd}, MDIVL \cite{db_mdivl}, MDID \cite{db_mdid}, MDID2013 \cite{md_sisblim_db_mdid2013}, and LIVE WCmp \cite{md_2stepQA_jrnl}, use a limited number of distortion levels (typically 3 or 4) per distortion type per stage, making it difficult to analyze how different constituent distortions behave in conjunction with each other. The Waterloo Exploration-II (Exp-II) database \cite{dnn_database_tip} has 3,570 PR, 117,810 singly distorted, and 3,337,950 multiply distorted images, offering an excellent testbed for two-stage DR IQA. Here we construct two new datasets, namely DR IQA database Version 1 (V1) and DR IQA database Version 2 (V2), by following the same procedure as \cite{dnn_database_tip}, but without any cross-dataset content overlap. The purpose here is to use one dataset for training and the other for validation in a machine learning process, and then use the Waterloo Exp-II database for independent testing.

A total of 68 pristine quality reference images were taken from the following sources: IQA databases CSIQ \cite{fr_mad_db_csiq}, IVC \cite{db_ivc}, LIVE R2 \cite{stateval_db_liveR2}, TID2013 \cite{db_tid2013}, Toyoma \cite{db_mict} and some pristine images were extracted from raw videos available at CDVL \cite{cdvl}. These images were divided into two disjoint groups of 34 images each, with one group forming the pristine image set of DR IQA database V1 and the other forming the pristine image set of DR IQA database V2.


\begin{table}[t!]
	\scriptsize
	\caption{Composition of DR IQA databases V1 and V2.}
	\vspace{-2mm}
	\centering
	\begin{tabular}{ c | c  c | c  c }
		\hline
		Reference Images & \multicolumn{2}{c |}{Stage-1 Distorted Images} & \multicolumn{2}{c}{Stage-2 Distorted Images} \\
		in each Database & \multicolumn{2}{c |}{in each Database} & \multicolumn{2}{c}{in each Database} \\
		(Pristine Quality) & \multicolumn{2}{c |}{(Singly Distorted DRs)} & \multicolumn{2}{c}{(Multiply Distorted FDs)} \\ \hline
		Number of & \multirow{2}{*}{Distortion} & Number of & Distortion & Number of \\ 
		Images & & Images & Combination & Images \\ \hline
		\multirow{7}{*}{34} & \multirow{2}{*}{Blur} & \multirow{2}{*}{374} & Blur-JPEG & 6,358 \\
		 & & & Blur-Noise & 6,358 \\ \cline{2-5}
		 & JPEG & 374 & JPEG-JPEG & 6,358 \\ \cline{2-5}
		 & \multirow{2}{*}{Noise} & \multirow{2}{*}{374} & Noise-JPEG & 6,358 \\
		 & & & Noise-JP2K & 6,358 \\ \cline{2-5}
		 & Total & 1,122 & Total & 31,790 \\ \cline{2-5}
		 & \multicolumn{4}{c}{Overall 32,912 Distorted Images in each Database} \\ \hline
	\end{tabular}
	\label{Table4_6}
	\vspace{-2mm}
\end{table}


\begin{figure}[t!]	
	\begin{minipage}[b]{0.5\textwidth}
		\centering
		\centerline{\includegraphics[width=0.67\linewidth]{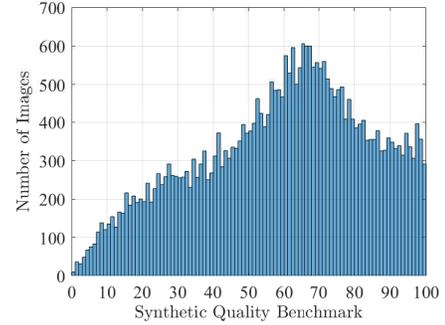}}
		\vspace{-1mm}
		\centerline{(a) DR IQA Database V1}\medskip
	\end{minipage}
	\hfill
	\begin{minipage}[b]{0.5\textwidth}
	    \vspace{-2mm}
		\centering
		\centerline{\includegraphics[width=0.67\linewidth]{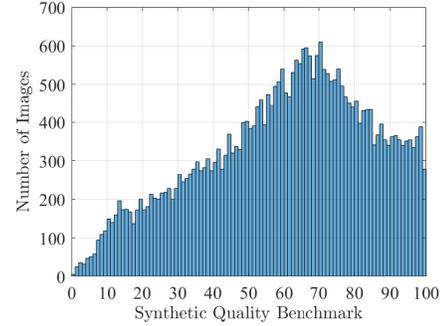}}
		\vspace{-1mm}
		\centerline{(b) DR IQA Database V2}\medskip
	\end{minipage}
	\vspace{-8mm}
	\caption{SQB histograms of the DR IQA databases V1 and V2.}
	\label{DRIQA_V12_SQBhist}
	\vspace{-4mm}
\end{figure}

Table \ref{Table4_6} outlines the composition of DR IQA databases V1 and V2. We include singly distorted DR images belonging to three distortion categories of Gaussian white noise, Gaussian blur, and JPEG compression. The stage-1 DR images are created in the \textit{fair} to \textit{excellent} perceptual quality range at distortion levels 1 to 11 based on a content adaptive distortion process~\cite{dnn_database_tip}. This leads to 374 DR images for each of the three single distortion types in each dataset. The stage-2 multiply distorted FD images belong to five distortion combinations of Blur-JPEG (B-JPG), Blur-Noise (B-N), JPEG-JPEG (JPG-JPG), Noise-JPEG (N-JPG), and Noise-JPEG2000 (Noise-JP2K or N-JP2), which represent real-world multiple distortion scenarios such as different image capture conditions, storage of noisy or blurry images through compression, and multiple levels of compression. This is done by starting with the respective DR images and applying content adaptive distortion parameters~\cite{dnn_database_tip} at levels 1 to 17, which correspond to the entire \textit{bad} to \textit{excellent} perceptual quality range. This leads to 6,358 FD images for each distortion combination in each dataset. By having 11 stage-1 and 17 stage-2 distortion levels, we have ensured an adequate density of distortion levels per distortion stage. Overall, each of the V1 and V2 databases has 32,912 distorted images. Conducting subjective testing for such large datasets is extremely difficult. To annotate large-scale IQA datasets without subjective testing, we conducted the largest IQA performance evaluation study to-date in~\cite{eval_Waterloo} based on which we developed an alternative mechanism for IQA data annotation called synthetic quality benchmark (SQB) in~\cite{dnn_database_tip}, where its efficacy was also established. We use SQB to annotate DR IQA databases V1 and V2, and provide their SQB histograms in Fig.~\ref{DRIQA_V12_SQBhist}, where both databases give a wide representation of the full quality range (SQB dynamic range is between 0 and 100, where the best quality is represented by the latter), with a higher concentration on the higher quality half, which is the working range in most real-world applications.


\begin{figure*}[t!]
	\centering
	\begin{tabular}{c c c c c}
		 &
		 &
		\includegraphics[width=0.175\textwidth]{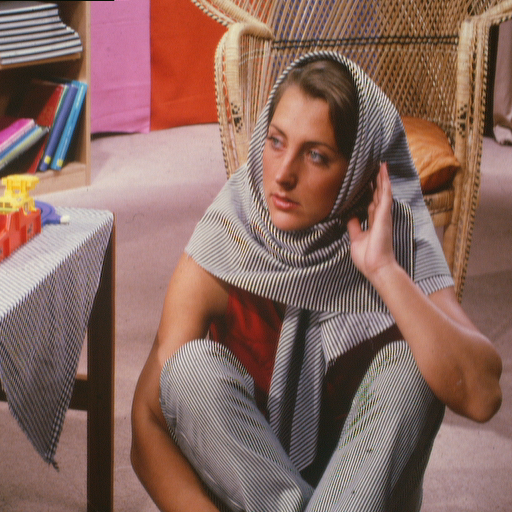} &
		 &
		 \\
		 &  & (a) PR &  &  \\[6pt]
	\end{tabular}
	\begin{tabular}{c c c c c}
		\multicolumn{2}{c}{\includegraphics[width=0.175\textwidth]{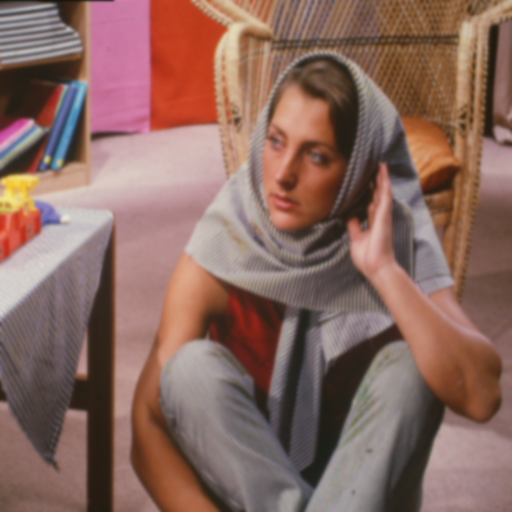}} &
		\includegraphics[width=0.175\textwidth]{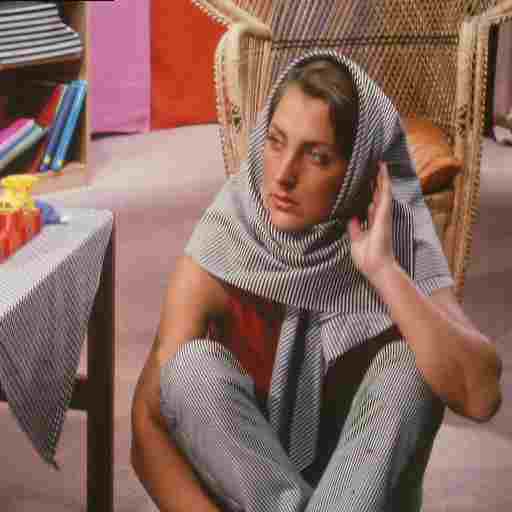} &
		\multicolumn{2}{c}{\includegraphics[width=0.175\textwidth]{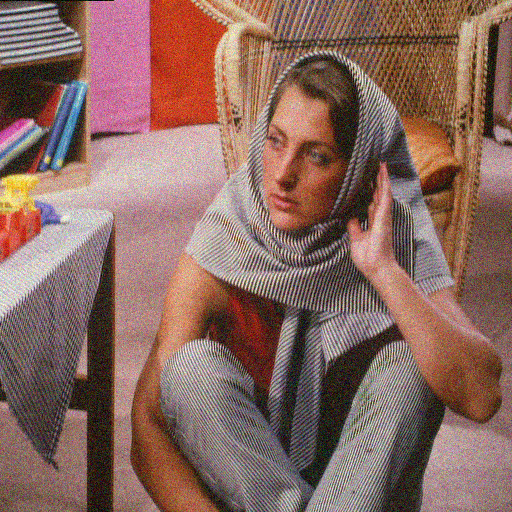}} \\
		\multicolumn{2}{c}{(b) DR (Blur)} & (c) DR (JPEG) & \multicolumn{2}{c}{(d) DR (Noise)} \\[6pt]
	\end{tabular}
	\begin{tabular}{c c c c c}
		\includegraphics[width=0.175\textwidth]{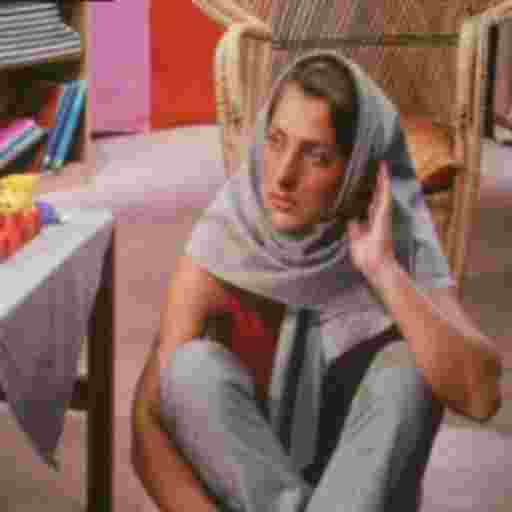} &
		\includegraphics[width=0.175\textwidth]{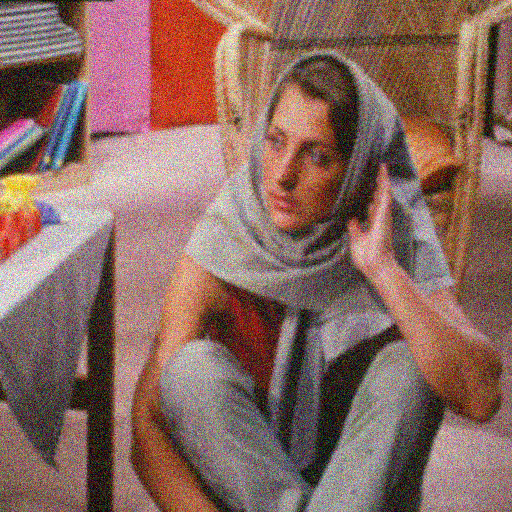} &
		\includegraphics[width=0.175\textwidth]{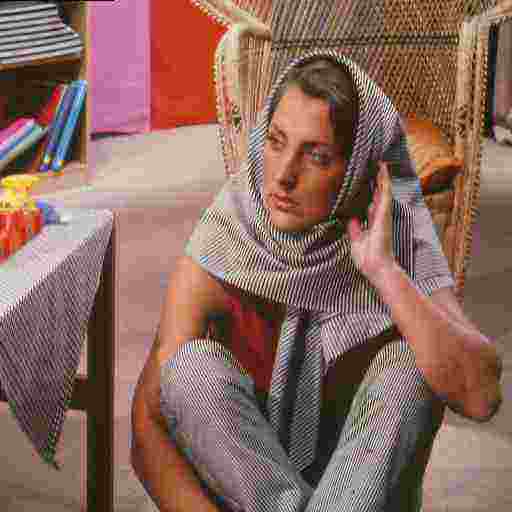} &
		\includegraphics[width=0.175\textwidth]{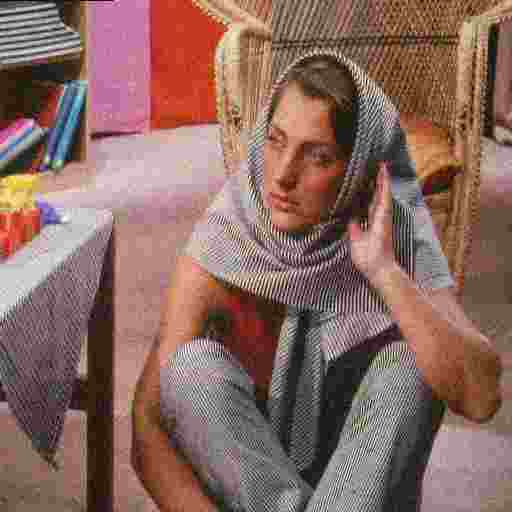} &
		\includegraphics[width=0.175\textwidth]{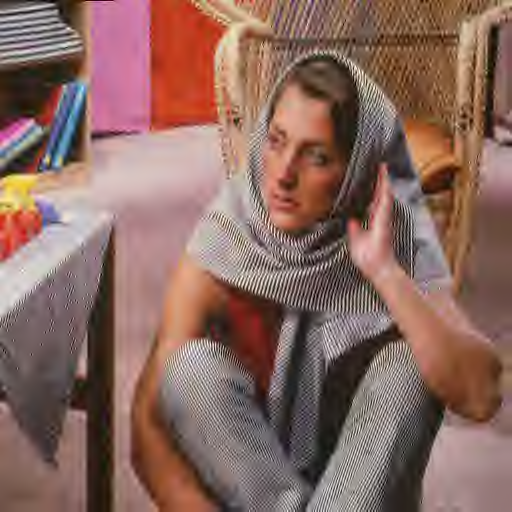} \\
		(e) FD (Blur-JPEG)  & (f) FD (Blur-Noise) & (g) FD (JPEG-JPEG) & (h) FD (Noise-JPEG) & (i) FD (Noise-JP2K) \\[6pt]
	\end{tabular}
	\vspace{-2mm}
	\caption{Example \textit{Barbara} images: (a) PR image; (b-d) DR images (Stage 1 distortion level 7); (e-i) FD images (Stage 2 distortion level 11).}
	\label{fig:Barbara_Images}
\end{figure*}


\begin{figure*}[t!]
	\centering
	\begin{tabular}{l c c c c c}
		\rotatebox{90}{\small{Absolute Quality Maps}} & \multicolumn{2}{c}{\includegraphics[width=0.175\textwidth]{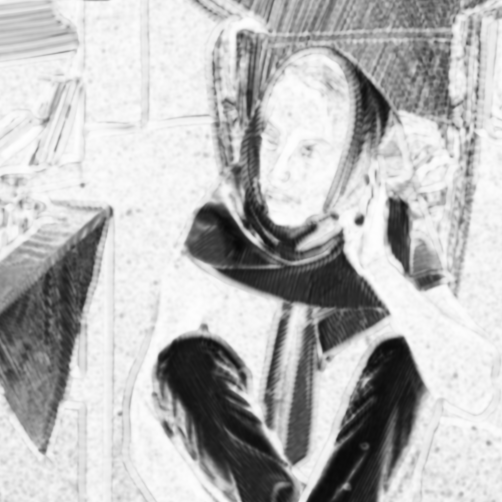}} &
		\includegraphics[width=0.175\textwidth]{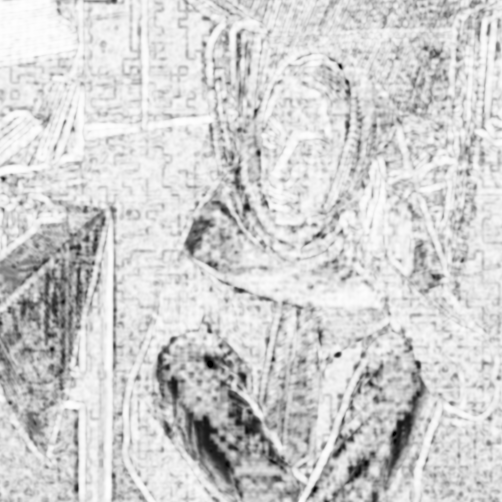} &
		\multicolumn{2}{c}{\includegraphics[width=0.175\textwidth]{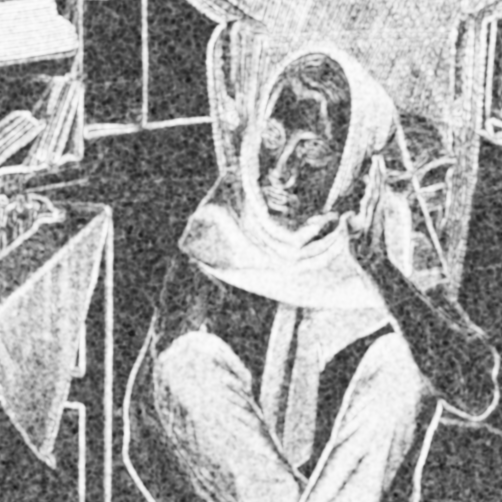}} \\
		& \multicolumn{2}{c}{(a) DR (Blur)} & (b) DR (JPEG) & \multicolumn{2}{c}{(c) DR (Noise)} \\[6pt]
	\end{tabular}
	\begin{tabular}{l c c c c c}
		\rotatebox{90}{\small{Relative Quality Maps}} &
		\includegraphics[width=0.175\textwidth]{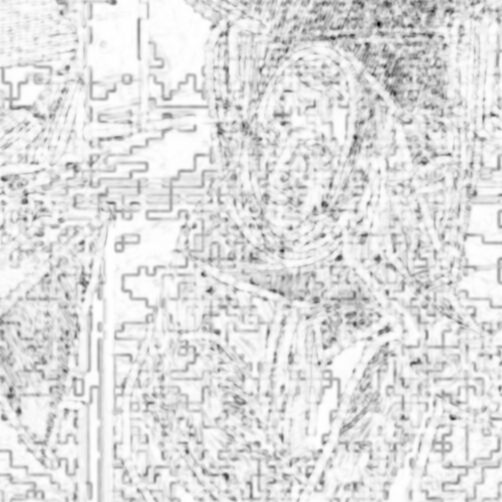} &
		\includegraphics[width=0.175\textwidth]{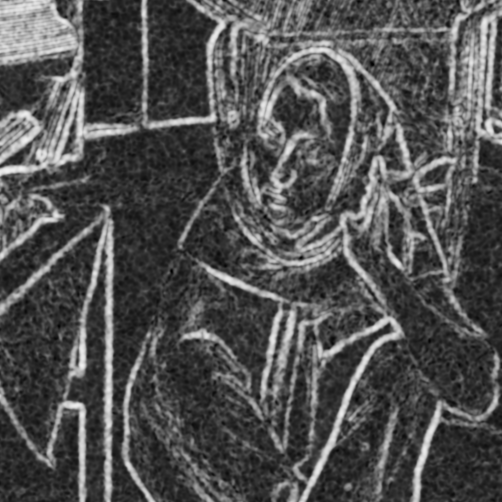} &
		\includegraphics[width=0.175\textwidth]{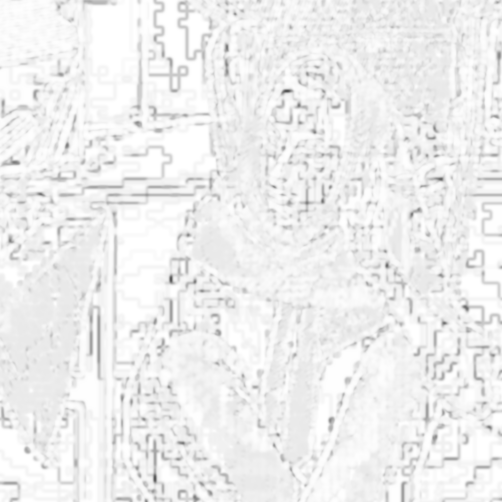} &
		\includegraphics[width=0.175\textwidth]{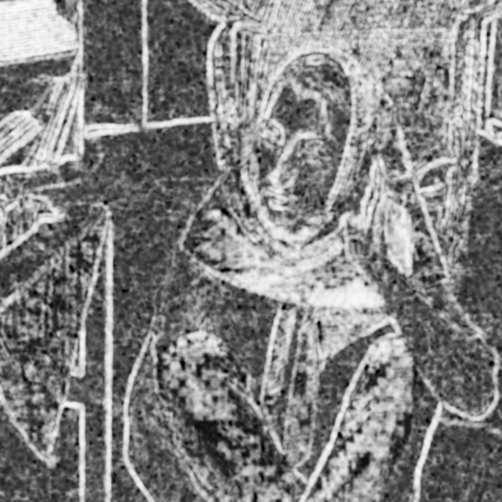} &
		\includegraphics[width=0.175\textwidth]{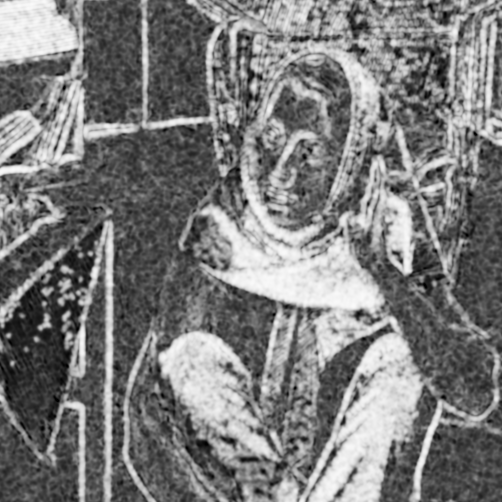} \\
		& (d) FD (Blur-JPEG)  & (e) FD (Blur-Noise) & (f) FD (JPEG-JPEG) & (g) FD (Noise-JPEG) & (h) FD (Noise-JP2K) \\[6pt]
	\end{tabular}
	\begin{tabular}{l c c c c c}
		\rotatebox{90}{\small{Absolute Quality Maps}} &
		\includegraphics[width=0.175\textwidth]{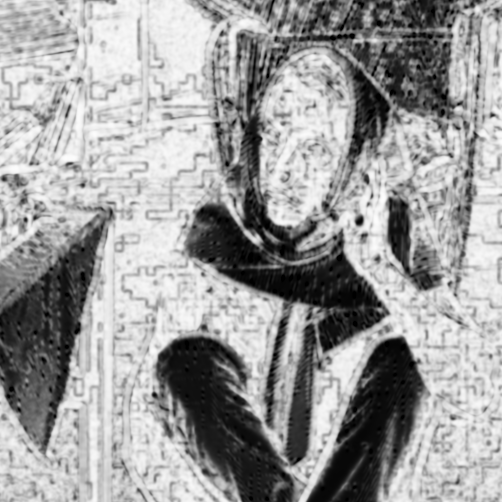} &
		\includegraphics[width=0.175\textwidth]{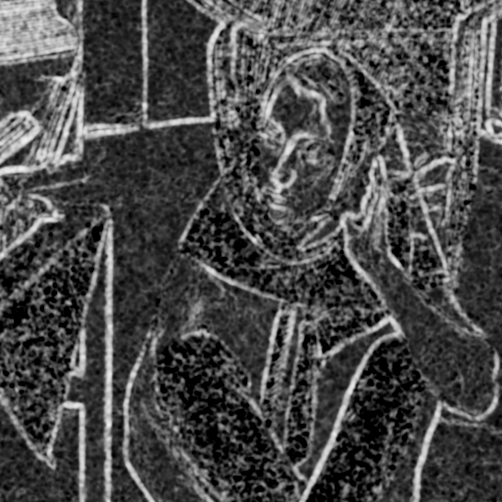} &
		\includegraphics[width=0.175\textwidth]{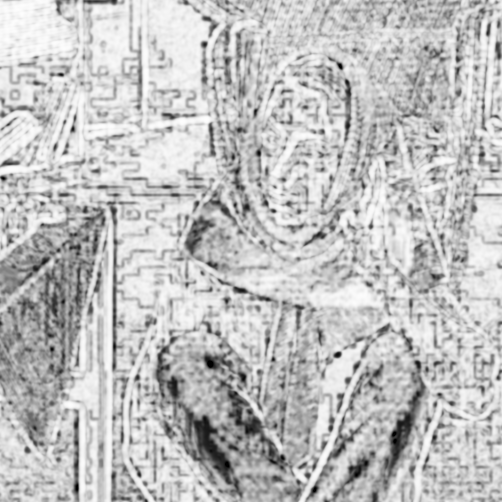} &
		\includegraphics[width=0.175\textwidth]{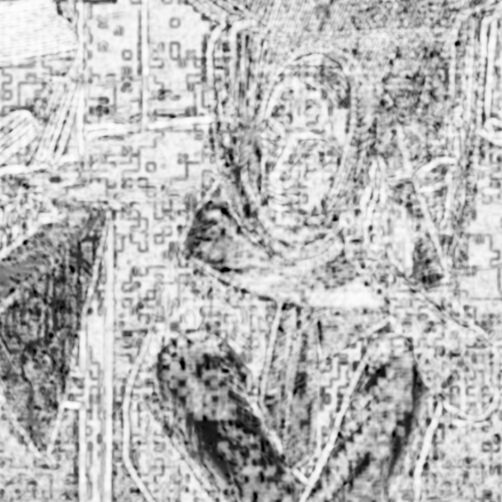} &
		\includegraphics[width=0.175\textwidth]{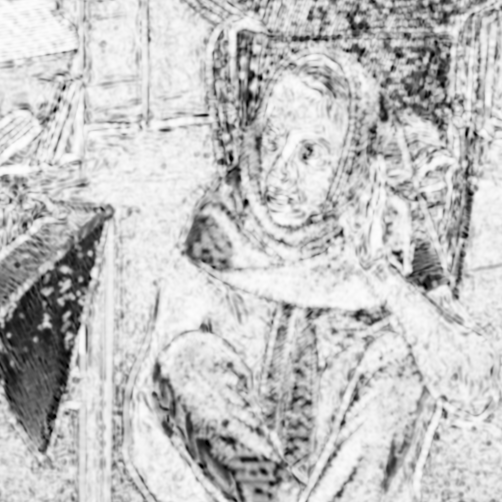} \\
		& (i) FD (Blur-JPEG)  & (j) FD (Blur-Noise) & (k) FD (JPEG-JPEG) & (l) FD (Noise-JPEG) & (m) FD (Noise-JP2K) \\[6pt]
	\end{tabular}
	\vspace{-4mm}
	\caption{SSIM \cite{fr_ssim} Quality Maps of the example \textit{Barbara} images of Fig. \ref{fig:Barbara_Images}: (a-c) Absolute Quality Maps of the DR images with respect to PR; (d-h) Relative Quality Maps of the FD images with respect to their DR images; (i-m) Absolute Quality Maps of the FD images with respect to PR.}
	\label{fig:QMaps}
	\vspace{-3mm}
\end{figure*}


\begin{table}[t!]
	\scriptsize
	\caption{FSIMc $\text{AS}_{\text{DR}}$, $\text{RS}_{\text{FD}}$, and $\text{AS}_{\text{FD}}$ scores for examples in Fig. \ref{fig:Barbara_Images}}
	\vspace{-3mm}
	\centering
	\begin{tabular}{c|c|c|c|c|c|c}
		\hline
		Distortion & \multicolumn{2}{c|}{FSIMc $\text{AS}_{\text{DR}}$} & \multicolumn{2}{c|}{FSIMc $\text{RS}_{\text{FD}}$} & \multicolumn{2}{c}{FSIMc $\text{AS}_{\text{FD}}$} \\ \cline{2-7}
		Combination & between & score & between & score & between & score \\ \hline
		Blur-JPEG & (a)\&(b) & 0.9495 & (b)\&(e) & 0.9253 & (a)\&(e) & 0.8815 \\ \hline
		Blur-Noise & (a)\&(b) & 0.9495 & (b)\&(f) & 0.8646 & (a)\&(f) & 0.8570 \\ \hline
		JPEG-JPEG & (a)\&(c) & 0.9525 & (c)\&(g) & 0.9538 & (a)\&(g) & 0.9066 \\ \hline
		Noise-JPEG & (a)\&(d) & 0.9311 & (d)\&(h) & 0.8969 & (a)\&(h) & 0.9016 \\ \hline
		Noise-JP2K & (a)\&(d) & 0.9311 & (d)\&(i) & 0.8907 & (a)\&(i) & 0.9148 \\ \hline
	\end{tabular}
	\label{Table4_5}
	\vspace{-5mm}
\end{table}


\begin{figure*}[t!]
	\centering
	\begin{tabular}{c c c}
		\includegraphics[width=0.28\textwidth]{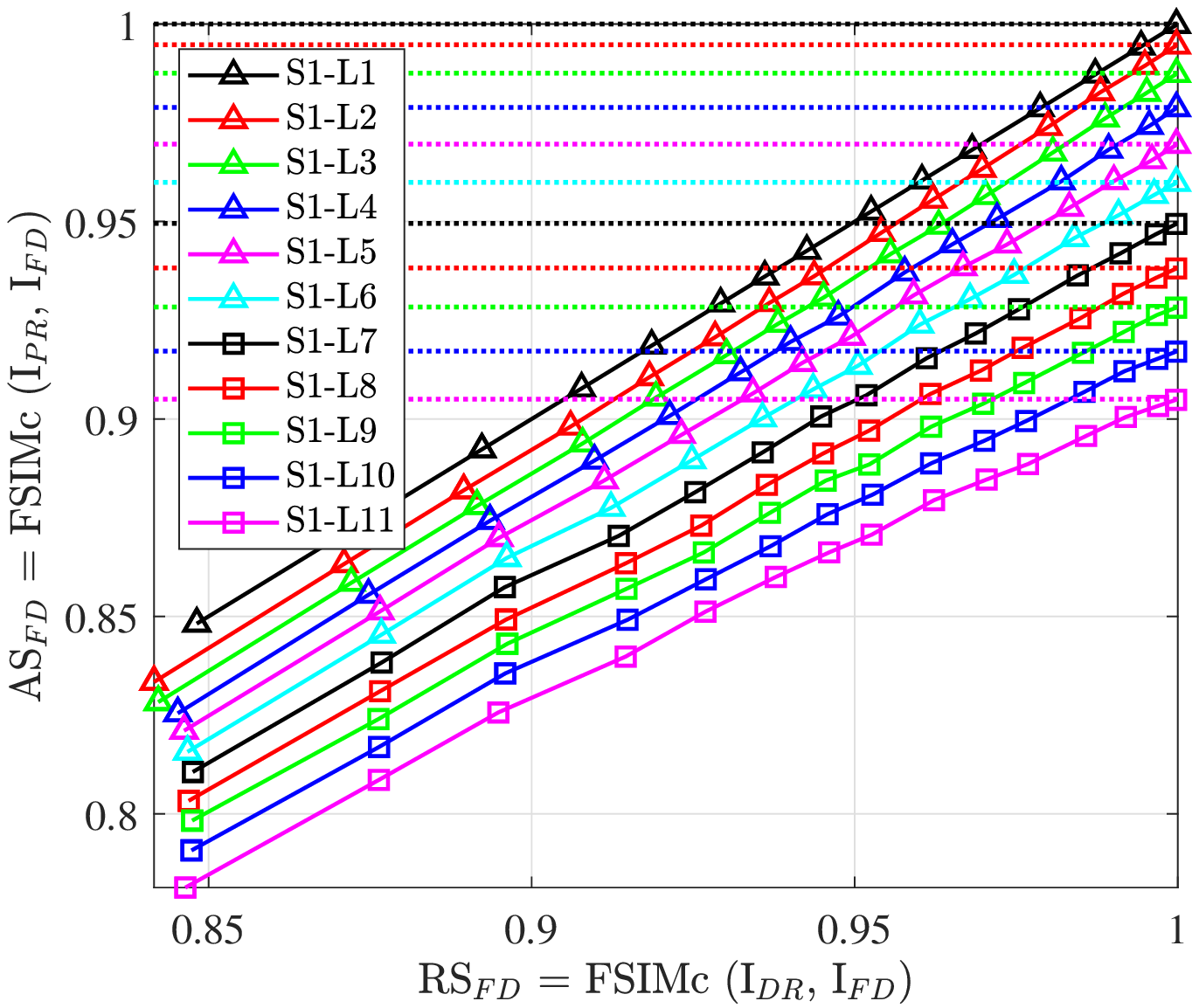} &
		\includegraphics[width=0.28\textwidth]{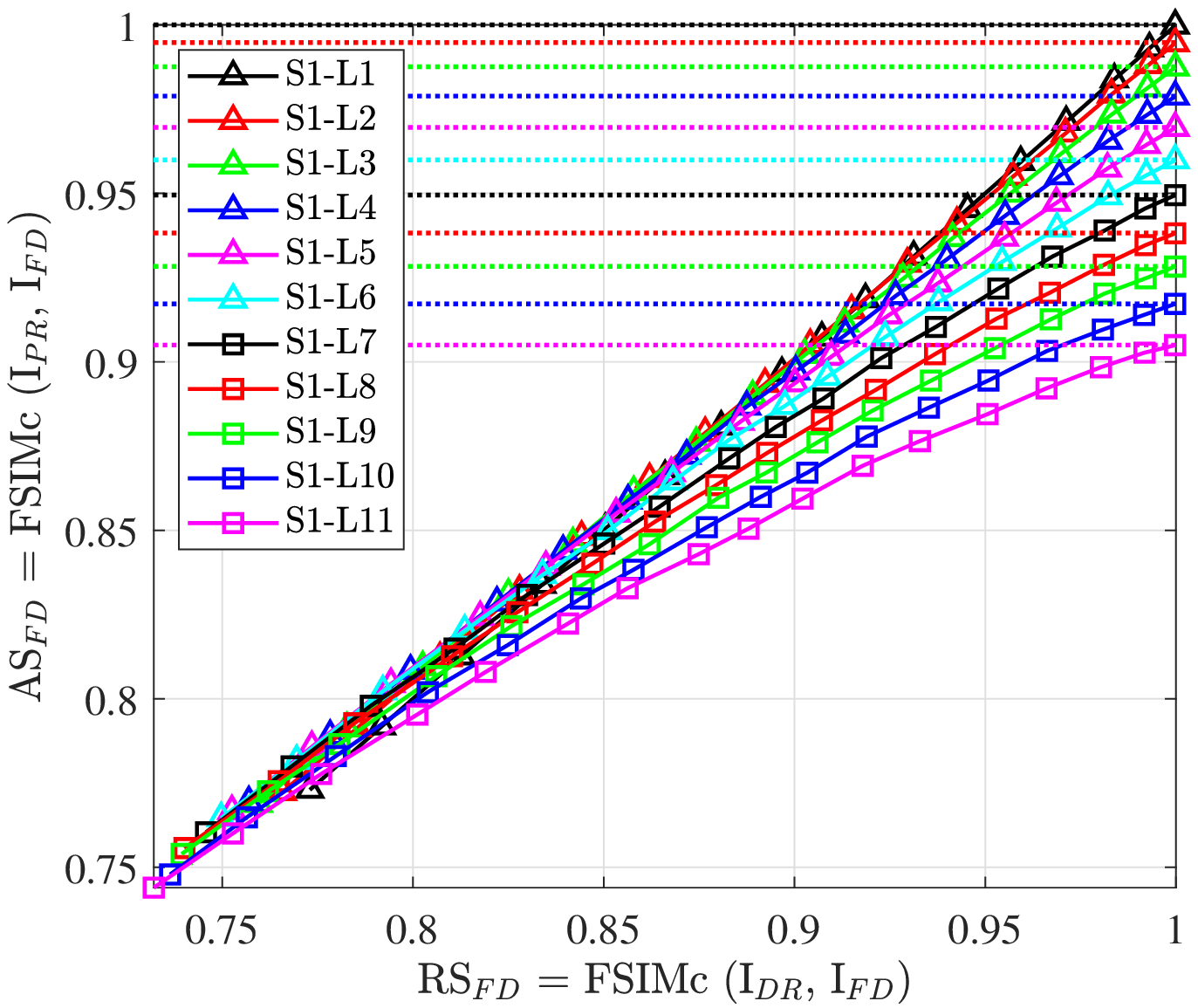} &
		\includegraphics[width=0.28\textwidth]{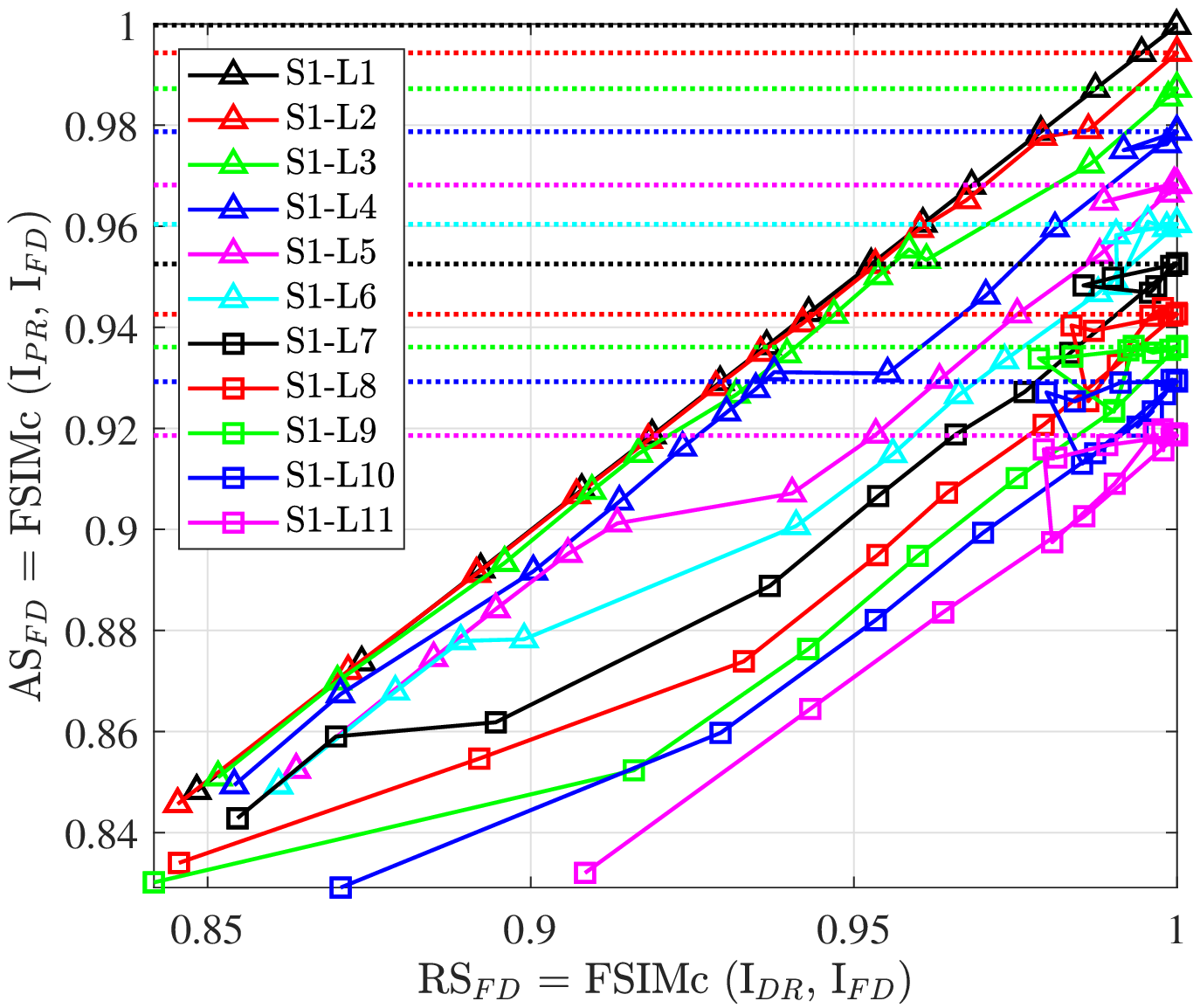} \\
		(a) Blur-JPEG  & (b) Blur-Noise & (c) JPEG-JPEG \\[6pt]
	\end{tabular}
	\begin{tabular}{c c}
		\includegraphics[width=0.28\textwidth]{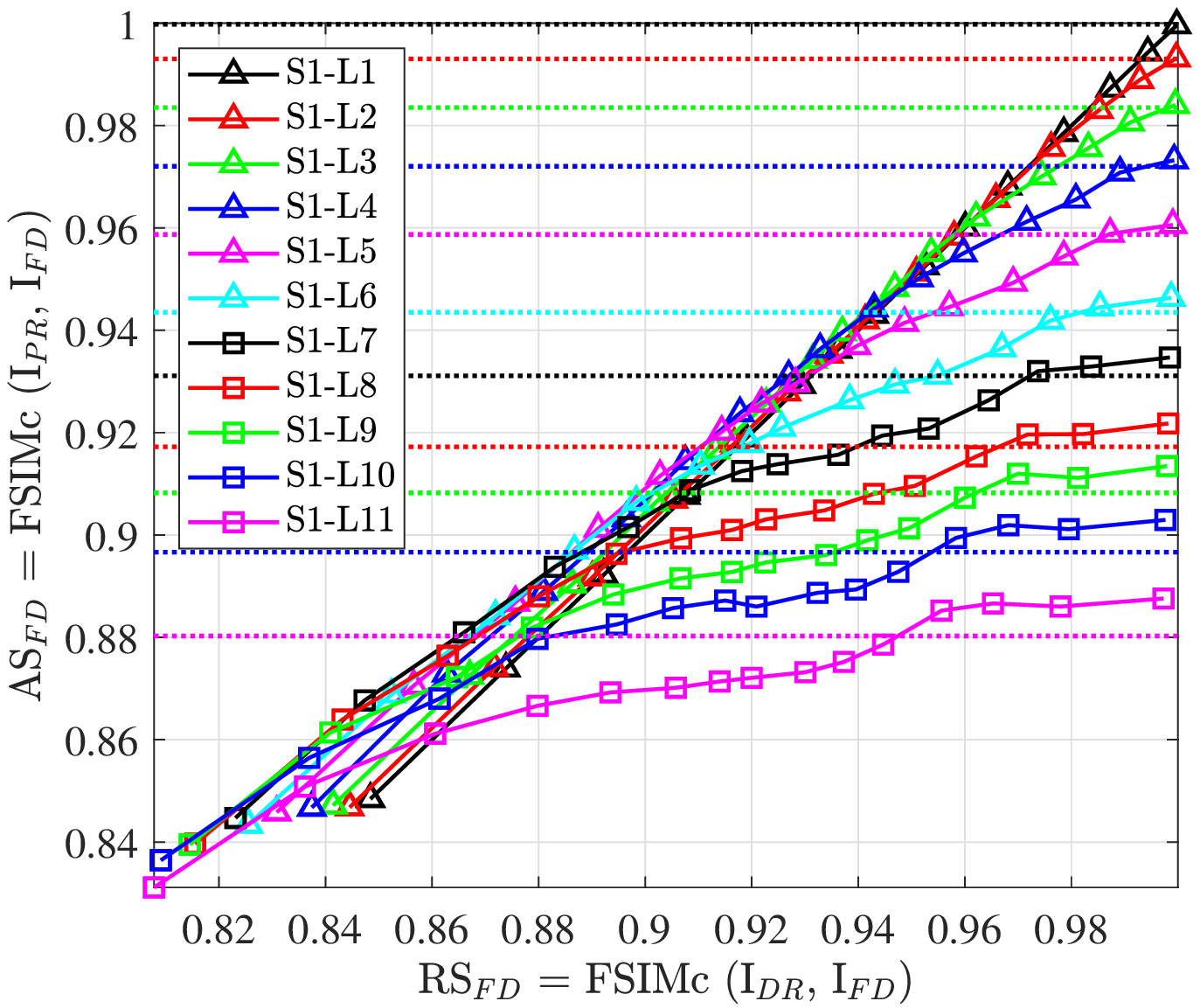} &
		\includegraphics[width=0.28\textwidth]{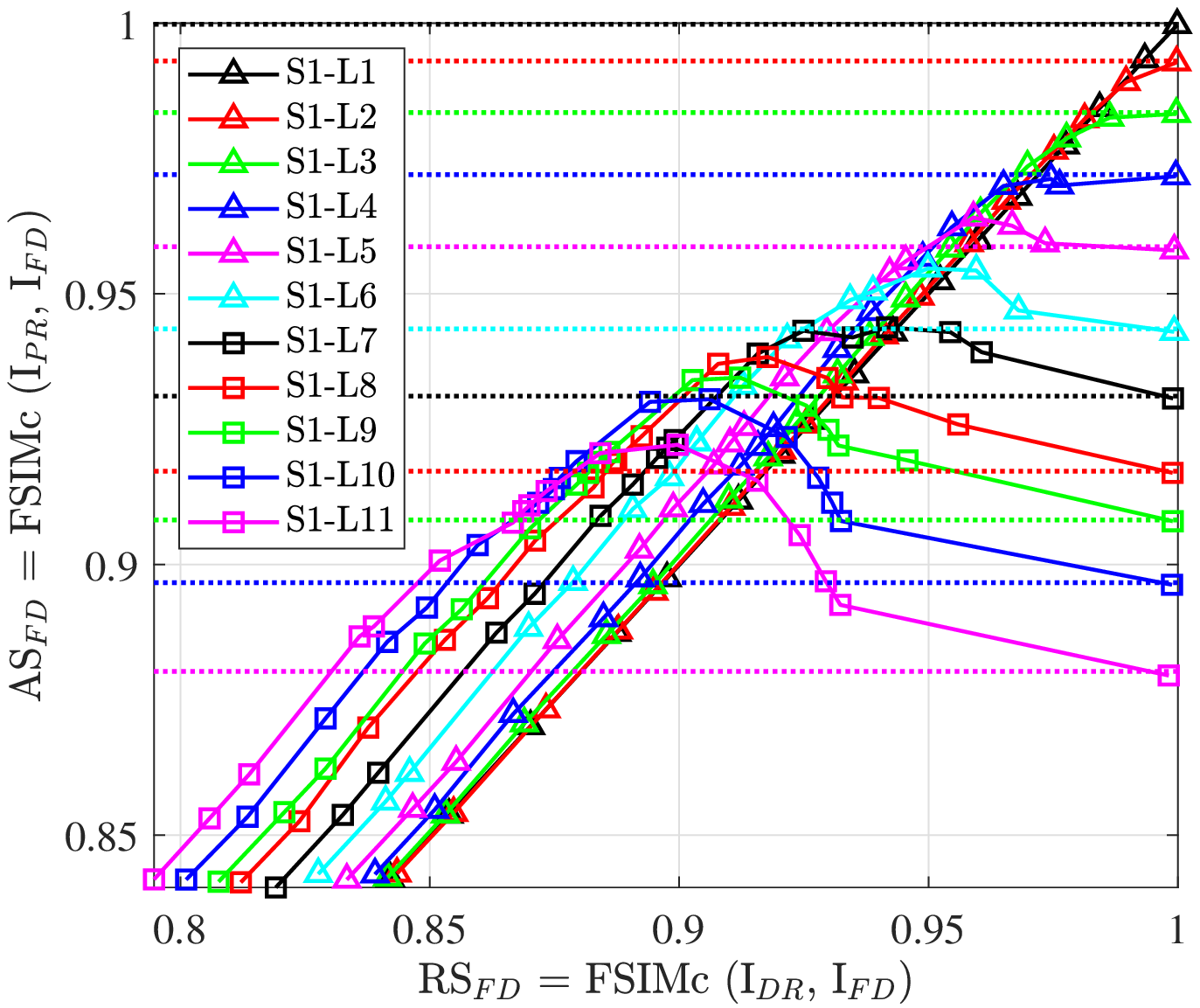} \\
		(d) Noise-JPEG  & (e) Noise-JP2K \\[6pt]
	\end{tabular}
	\vspace{-3mm}
	\caption{$\text{AS}_{\text{FD}}$ versus $\text{RS}_{\text{FD}}$ plots for all stage-1 ($\text{AS}_{\text{DR}}$) distortion levels (SL-L1 to SL-L11) corresponding to five distortion combination types for the \textit{Barbara} image. Dotted lines represent $\text{AS}_{\text{DR}}$ scores.}
	\label{fig:Plots_ASfd_vs_RSfd}
	\vspace{-5mm}
\end{figure*}

\subsection{Multiple Distortions: Behavior Analysis}
\label{sec:MDbehaveAnalysis}

The large-scale DR IQA databases V1 and V2 provide us with a platform to investigate the behaviors of images undergoing multiple stages of distortions, which is essential in building effective DR IQA methods. We start with a visual example given in Figs.~\ref{fig:Barbara_Images} and \ref{fig:QMaps}, where the PR \textit{Barbara} image undergoes three types of stage-1 distortions (Gaussian noise, Gaussian blur, and JPEG compression) at level 7, which then undergo stage-2 distortions (Gaussian noise, JPEG and JPEG2000 compression) at level 11 to create five distortion combinations (Blur-JPEG, Blur-Noise, JPEG-JPEG, Noise-JPEG, Noise-JP2K), resulting in 3 DR and 5 FD images, for which we compute their SSIM~\cite{fr_ssim} quality maps that indicate quality variations over space (brighter suggests better quality). The various image-level $\text{AS}_{\text{DR}}$, $\text{AS}_{\text{FD}}$, and $\text{RS}_{\text{FD}}$ quality scores for these images are computed by the FR IQA method FSIMc~\cite{fr_fsim} and are given in Table~\ref{Table4_5}.

There are several useful observations from these visual examples and FSIMc results. 1) The relative quality maps of the FD images (Fig. \ref{fig:QMaps}(d-h)) are drastically different from their respective absolute quality maps (Fig. \ref{fig:QMaps}(i-m)), suggesting $\text{RS}_{\text{FD}}$ is not a good predictor of $\text{AS}_{\text{FD}}$ in general and FR methods should be used with caution in the absence of PR images. 2) For the cases of Blur-JPEG, Blur-Noise, and JPEG-JPEG, the relative quality maps are lighter than the absolute quality maps, showing an over-estimation of $\text{AS}_{\text{FD}}$ by $\text{RS}_{\text{FD}}$. 3) However, the opposite is observed for the cases of Noise-JPEG and Noise-JP2K, where their relative quality maps are darker, and $\text{RS}_{\text{FD}}$ under-estimates $\text{AS}_{\text{FD}}$. 4) For the cases of Blur-JPEG, Blur-Noise, and JPEG-JPEG, the absolute quality map of the DR image and the relative quality map of the FD image appear to be roughly additive with regard to the absolute quality map of the FD image. 5) However, this is obviously not the case for Noise-JPEG and Noise-JP2K, suggesting sophisticated distortion combination behaviors.

To investigate further, we use the FR method FSIMc \cite{fr_fsim} to compute the $\text{AS}_{\text{DR}}$, $\text{RS}_{\text{FD}}$, and $\text{AS}_{\text{FD}}$ scores for various DR and FD \textit{Barbara} images created from all 11-level stage-1 distortions and 17-level stage-2 distortion combinations, and plot $\text{AS}_{\text{FD}}$ versus $\text{RS}_{\text{FD}}$ scores in Fig. \ref{fig:Plots_ASfd_vs_RSfd}. For each distortion combination, there are 11 curves, each corresponding to one of the 11 stage-1 DR images and containing 17 points that represent the corresponding stage-2 FD images. The dotted horizontal lines represent the quality level of the DR images.

\begin{figure}[t!]
	\centering
	\begin{tabular}{c c}
		\includegraphics[width=0.175\textwidth]{Figures/i04.png} &
		\includegraphics[width=0.175\textwidth]{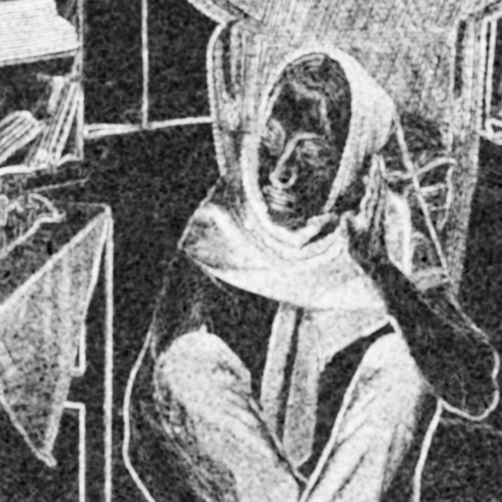} \\	
		(a) & (d) \\[4pt]
	\end{tabular}
	\begin{tabular}{c c}
		\includegraphics[width=0.175\textwidth]{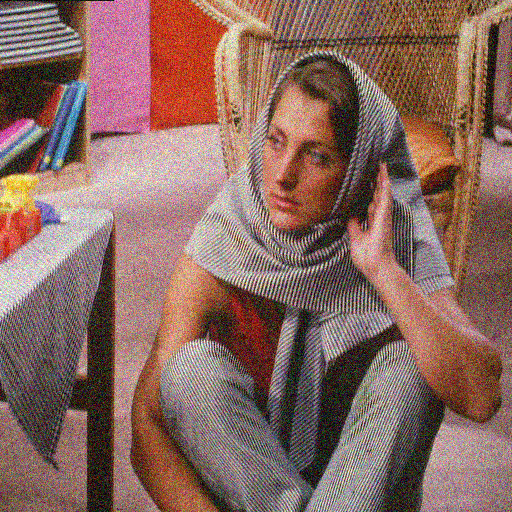} &
		\includegraphics[width=0.175\textwidth]{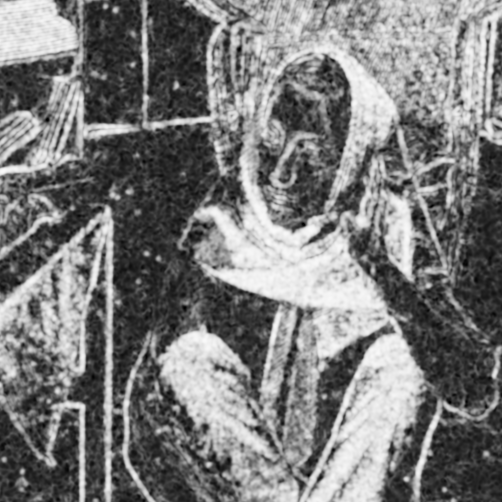} \\
		(b) & (e) \\[4pt]
	\end{tabular}
	\begin{tabular}{c c}
		 \includegraphics[width=0.175\textwidth]{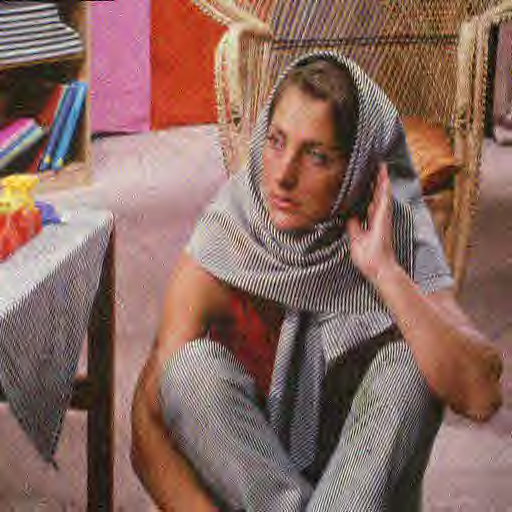} &
		 \includegraphics[width=0.175\textwidth]{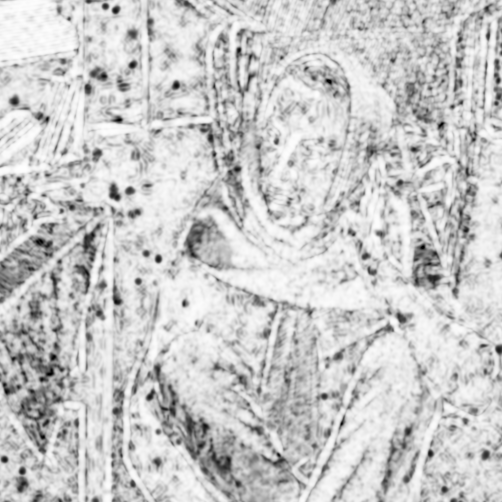} \\
		(c) & (f) \\[4pt]
	\end{tabular}
	\vspace{-2mm}
	\caption{Noise-JP2K distortion combination example. (a) PR \textit{Barbara} image; (b) DR image contaminated by white Gaussian noise (level 11), FSIMc $\text{AS}_{\text{DR}}$ = 0.8802; (c) FD image obtained by compressing the DR image with JPEG2000 (level 6), FSIMc $\text{RS}_{\text{FD}}$ = 0.8997, FSIMc $\text{AS}_{\text{FD}}$ = 0.9221; SSIM quality maps of: (d) DR with respect to PR image; (e) FD with respect to DR image; (f) FD with respect to PR image.}
	\label{fig:Barbara_NJP2_Ex2}
	\vspace{-6mm}
\end{figure}

The key observations are summarized as follows: 1) When stage-1 distortion is minimum (level-1), the DR image is almost as good as the PR image, and not surprisingly the prediction from $\text{RS}_{\text{FD}}$ to $\text{AS}_{\text{FD}}$ is nearly perfect, demonstrated as the straight lines of the Stage-1 Level-1 (S1-L1) curves in all five distortion combinations in Fig.~\ref{fig:Plots_ASfd_vs_RSfd}. However, the trend changes dramatically with the increasing stage-1 distortion levels; 2) For all five distortion combinations, with the exception of some Noise-JPEG curves, at minimum stage-2 distortion (Stage-2 Level-1), $\text{RS}_{\text{FD}} \simeq 1$ and $\text{AS}_{\text{FD}} \simeq \text{AS}_{\text{DR}}$, which is not surprising since stage-2 is not adding any further distortion to the DR image; 3) For Blur-JPEG and JPEG-JPEG distortion combinations (Fig. \ref{fig:Plots_ASfd_vs_RSfd} (a) and (c)), the curves consistently move away nearly in parallel (especially the Blur-JPEG case) from the S1-L1 curve with increasing stage-1 distortion level, suggesting additive quality degradation of the two distortion stages; 4) The Blur-Noise distortion combination (Fig.~\ref{fig:Plots_ASfd_vs_RSfd}(b)) follows a similar behavior as the Blur-JPEG and JPEG-JPEG cases, but the curves converge with increasing stage-2 distortion, implying that as the magnitude of the stage-2 distortion, i.e., Gaussian noise, increases, it overshadows the stage-1 distortion (Gaussian blur) and becomes the dominant distortion factor; 5) The Noise-JPEG and Noise-JP2K distortion combination cases (Fig.~\ref{fig:Plots_ASfd_vs_RSfd} (d) and (e)) exhibit more interesting nonlinear behaviors. Notably some portions of the curves go above their respective $\text{AS}_{\text{DR}}$ baseline, i.e., overshoots take place. This behavior is most apparent in the low to mid-level stage-2 distortion levels corresponding to mid to high level stage-1 distortion levels and is much more pronounced in the Noise-JP2K case. Most interestingly, the overshoot phenomenon indicates that the corresponding FD images have better quality than their respective DR images from which they are created, suggesting that the denoising effect of compression may help improve image quality at certain noise and compression levels. A visual demonstration is given in Fig.~\ref{fig:Barbara_NJP2_Ex2}, where the PR \textit{Barbara} image is distorted at Gaussian noise level 11 to generate the DR image, which is then further distorted by JPEG2000 compression level 6 to generate the FD image. The quality maps of the DR and FD images with respect to the PR image and their FSIMc scores clearly show that the quality of the FD image improves upon the DR image. For Noise-JPEG and Noise-JP2K, especially the latter, we also note that unlike the other three distortion combinations, curves corresponding to higher stage-1 distortion levels are not always below those with lower stage-1 distortion. Instead as stage-2 distortion increases, a crossover takes places which is more evident for higher stage-1 distortion levels. Again this points to complex joint effects of the two distortion types involved in the combination where noisier stage-1 images are better impacted by the denosing effect of compression. Overall, the multi-stage quality variation behavior is highly dependent on the distortion combinations, and the relationship of the two distortion stages on quality ranges from nearly linear to highly nonlinear. While we have presented the above behavior analysis using one example due to space limit, similar behaviors have been observed in the whole DR IQA V1 and V2, and Waterloo Exp-II~\cite{dnn_database_tip} databases.


\section{DR IQA Model Design}
\label{sec:DRIQAmodeling}


\begin{figure*}[t!]
	\centering
	\begin{tabular}{c c c}
		\includegraphics[width=0.3\textwidth]{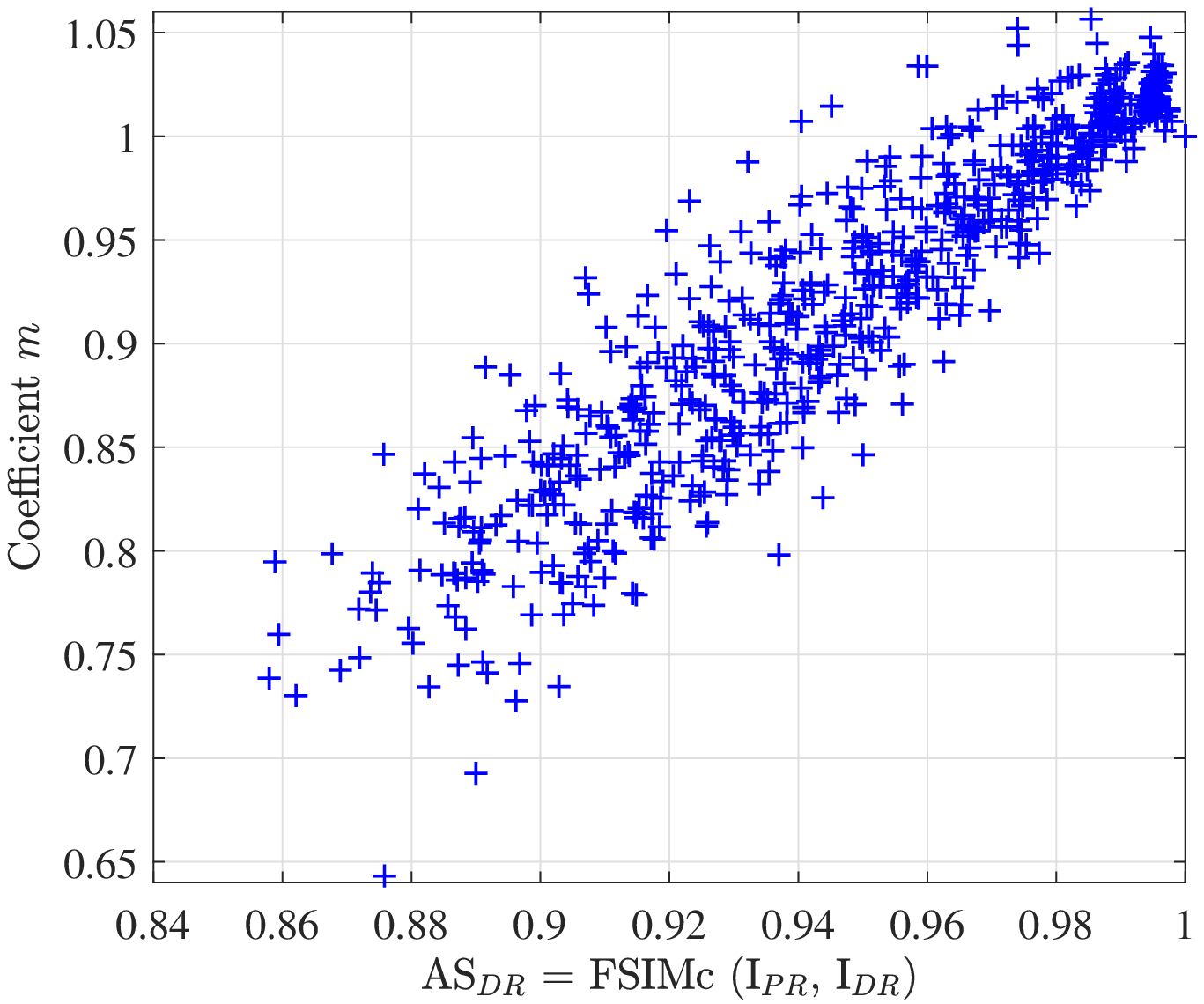} &
		\includegraphics[width=0.3\textwidth]{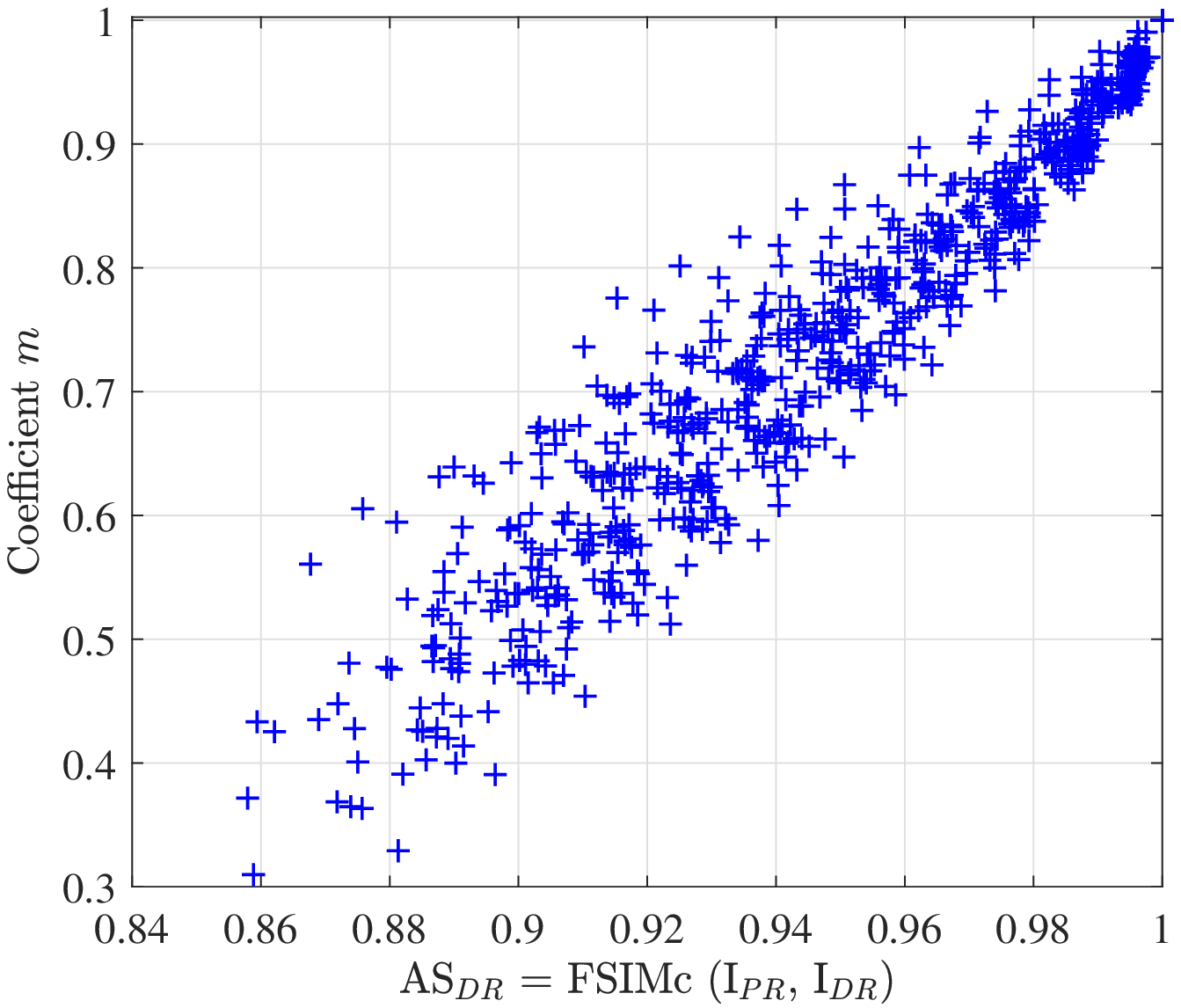} &
		\includegraphics[width=0.3\textwidth]{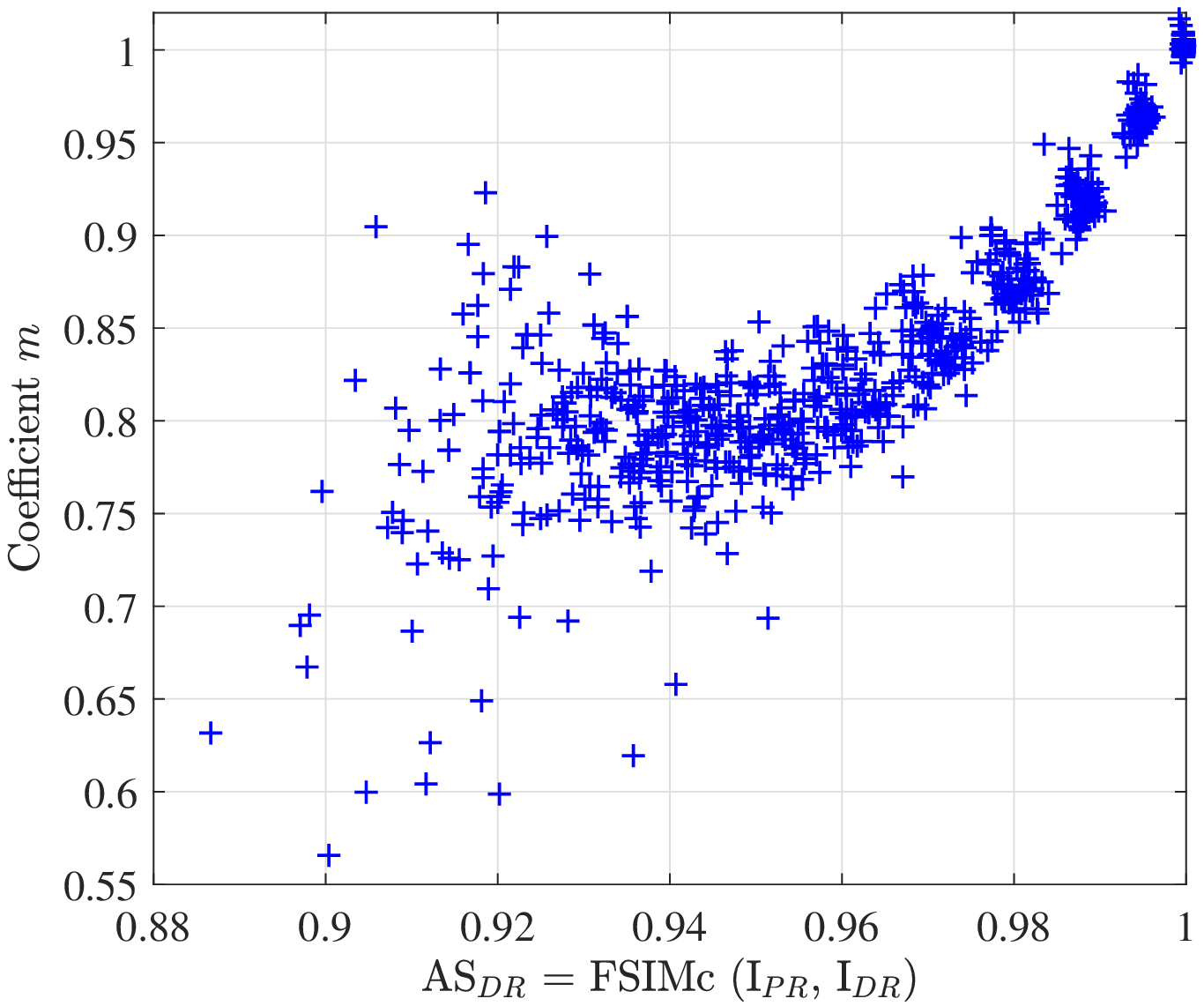} \\
		(a) Blur-JPEG  & (b) Blur-Noise & (c) JPEG-JPEG \\[6pt]
	\end{tabular}
	\begin{tabular}{c c}
		\includegraphics[width=0.3\textwidth]{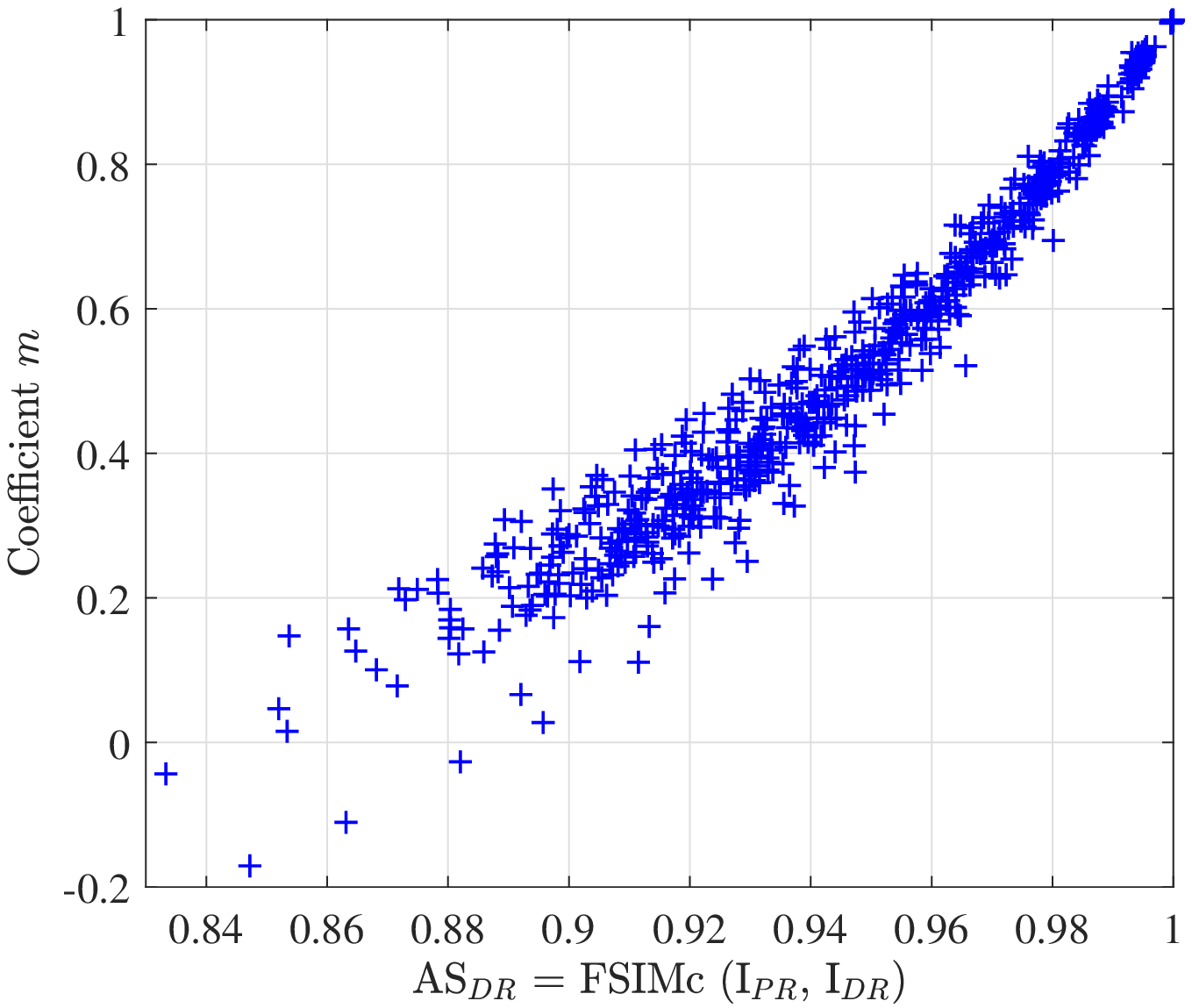} &
		\includegraphics[width=0.3\textwidth]{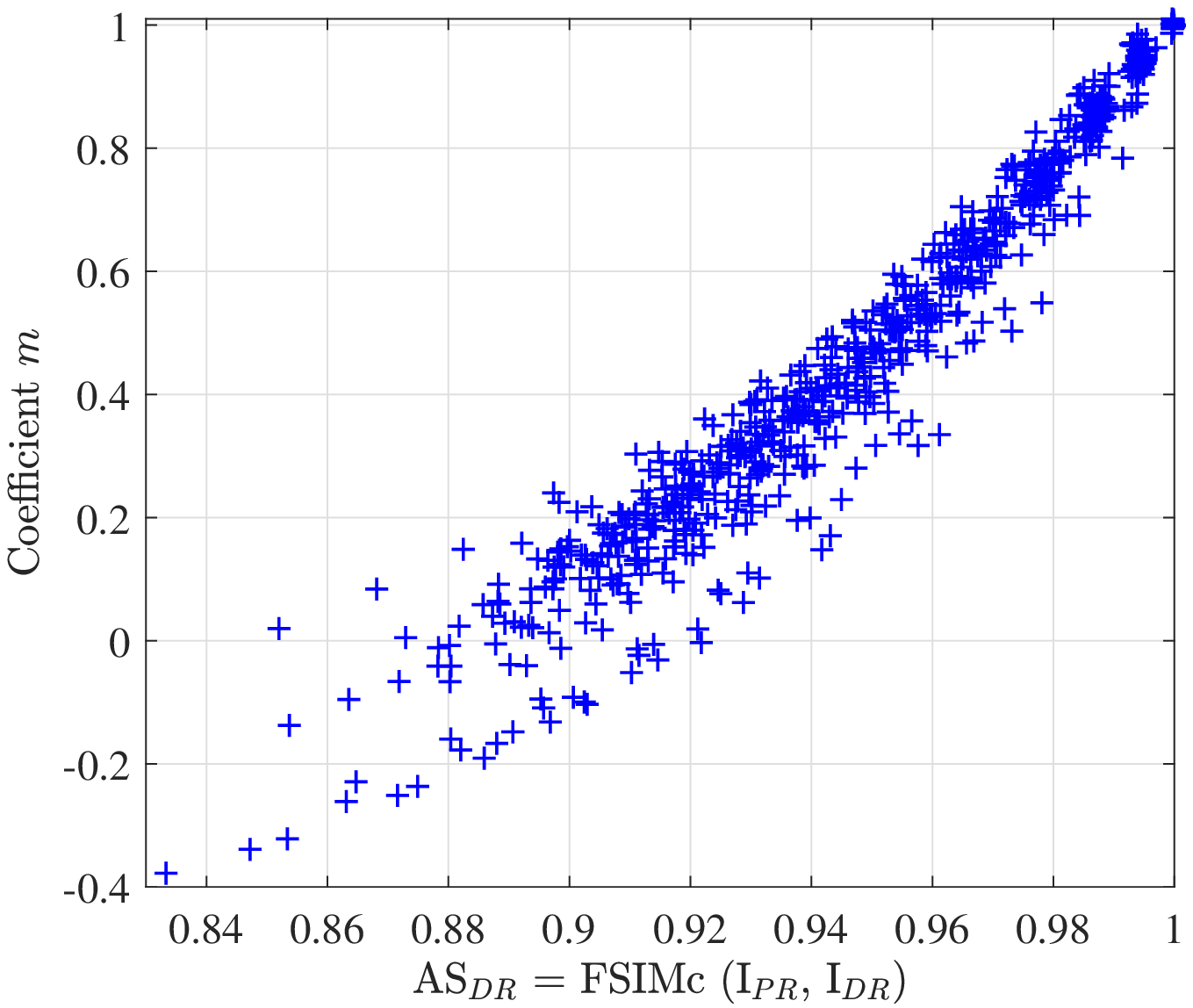} \\
		(d) Noise-JPEG  & (e) Noise-JP2K \\[6pt]
	\end{tabular}
	\vspace{-2mm}
	\caption{Scatter plots of coefficient $m$ versus $\text{AS}_{\text{DR}}$ for the entire DR IQA databases V1 and V2 for the five distortion combinations under consideration.}
	\label{fig:Coeff_m_AS-DR}
	\vspace{-4mm}
\end{figure*}

Here we make one of the first attempts to develop DR IQA models. Rather than adopting sophisticated methods such as deep neural networks, we opt for straightforward and empirical approaches, because our main goal is to establish DR IQA as a new paradigm that offers advantages in handling multiple-distortion cases through head-to-head comparisons with top-performing FR or NR methods (baselines in Section~\ref{sec:PerfEval}). The transparency of these approaches also demonstrates the value of understanding the multiple-distortion behaviors (as discussed in Section~\ref{sec:MDbehaveAnalysis}) in DR IQA modeling. 

\subsection{Model 1: Distortion Behavior Model}
\label{DRIQA_Model1}
The first model is motivated by the observations on distortion behaviors from Fig.~\ref{fig:Plots_ASfd_vs_RSfd}, where for each given stage-1 distortion level, the relationship between $\text{AS}_{\text{FD}}$ and $\text{RS}_{\text{FD}}$ is often well fitted by a straight line (with the exception of Noise-JP2K), anchored by the rightmost point at ($\text{RS}_{\text{FD}}$, $\text{AS}_{\text{FD}}$) = (1, $\text{AS}_{\text{DR}}$) for each curve, especially in Blue-JPEG, Blue-Noise, JPEG-JPEG, and Noise-JP2K cases. In Scenario 1, or Type-100100 DR IQA, shown in Fig.~\ref{DRIQA_Scenario1} and represented by Eq. \ref{DRIQA_Sc1}, both FR computed $\text{AS}_{\text{DR}}$ and $\text{RS}_{\text{FD}}$ are known, and thus an estimate of $\text{AS}_{\text{FD}}$ directly follows from the point-slope formula for each curve, given by

\begin{equation}
	\centering
	{\widehat{\text{AS}_{\text{FD}}} = m \cdot (\text{RS}_{\text{FD}}-1) + \text{AS}_{\text{DR}}.
\label{mod1-orig}}
\end{equation}

It remains to determine the slope parameter $m$. For each distortion combination, we use least-square regression to obtain the best value of $m$ for each of the 11 stage-1 distortion levels for all the DR images in DR-IQA databases V1 and V2. The plots of the best coefficient $m$ versus $\text{AS}_{\text{DR}}$ are shown in Fig.~\ref{fig:Coeff_m_AS-DR}. Somewhat surprisingly, for each distortion combination type (with the exception of JPEG-JPEG at high stage-1 distortion), the behavior of coefficient $m$ with respect to $\text{AS}_{\text{DR}}$ is rather quite linear, suggesting that the optimal value of $m$ may be directly predicted from $\text{AS}_{\text{DR}}$ by
\begin{equation}
	\centering
	{\widehat{m} = P_1 \cdot \text{AS}_{\text{DR}} + P_2,
		\label{submod}}
\end{equation}
where $P_1$ and $P_2$ are the slope and intercept coefficients, respectively. Replacing $m$ with $\widehat{m}$ and plugging into Eq.~\ref{mod1-orig}, we obtain
\begin{equation}
	\centering
	\scalebox{0.97}[1]{$\widehat{\text{AS}_{\text{FD}}} = P_1 \cdot \text{AS}_{\text{DR}} \cdot  \text{RS}_{\text{FD}} + P_2 \cdot \text{RS}_{\text{FD}} + (1-P_1) \cdot \text{AS}_{\text{DR}} - P_2.$}
		\label{mod1-modified3}
\end{equation}
As such, by following a 2-tier modeling approach, we narrow down the number of parameters to two ($P_1$ and $P_2$) for each distortion combination.

In addition to developing the 2-tier Model 1 above for the five distortion combinations separately, we also construct it for two other cases: 1) NBJ-JPEG (NBJ-JPG), where the distortion combinations of Noise-JPEG, Blur-JPEG, and JPEG-JPEG are considered together (so that comparisons can be made with 2stepQA \cite{md_2stepQA_conf,md_2stepQA_jrnl}, which is designed for the case where the second distortion stage is JPEG compression); and 2) the \textit{All Data} case, where all five distortion combinations are considered together. For Scenario 1, we use the FR method FSIMc \cite{fr_fsim} to compute both $\text{AS}_{\text{DR}}$ and $\text{RS}_{\text{FD}}$, and call this $\text{AS}_{\text{DR}}$-$\text{RS}_{\text{FD}}$ combination FSIMc-FSIMc.

For Scenario 2 or Type-001100 DR IQA as shown in Fig. \ref{DRIQA_Scenario2} and represented by Eq. \ref{DRIQA_Sc2}, we use three NR IQA methods, CORNIA \cite{nr_cornia}, dipIQ \cite{nr_dipiq}, and NIQE \cite{nr_niqe} to predict the quality of DR images, i.e., to find $\widehat{\text{AS}_{\text{DR}}}$ (Eq. \ref{AS_DRest}). CORNIA and dipIQ are selected for their top performance in \cite{eval_Waterloo}, and NIQE has been used in the 2stepQA model \cite{md_2stepQA_conf,md_2stepQA_jrnl}. Combining with the FR method FSIMc \cite{fr_fsim} (used to find $\text{RS}_{\text{FD}}$), this leads to three $\widehat{\text{AS}_{\text{DR}}}$-$\text{RS}_{\text{FD}}$ combinations: CORNIA-FSIMc, dipIQ-FSIMc, and NIQE-FSIMc. We also include the NIQE-MSSSIM combination, so as to make direct comparisons with 2stepQA \cite{md_2stepQA_conf,md_2stepQA_jrnl}. Specifically, we learn a nonlinear mapping from CORNIA, dipIQ, and NIQE to FSIMc for the CORNIA-FSIMc, dipIQ-FSIMc, and NIQE-FSIMc combinations, and from NIQE to MSSSIM for the NIQE-MSSSIM combination, with a five-parameter modified logistic function \cite{stateval_db_liveR2}:
\begin{equation}
	\centering
	{F(N) = \beta_1\Bigg[\frac{1}{2} - \frac{1}{1+e^{\{\beta_2(N-\beta_3)\}}}\Bigg] + \beta_4 N +\beta_5,
		\label{nlmapping_ch4}}
\end{equation}
where $N$ denotes NR scores, $F$ denotes mapped FR scores after the mapping step, and $\beta_1$, $\beta_2$, $\beta_3$, $\beta_4$, and $\beta_5$ are parameters tuned using DR IQA databases V1 and V2, and fixed for testing. The NR-predicted and FR-mapped $\widehat{\text{AS}_{\text{DR}}}$ score is then used in Model 1 (Eq. \ref{mod1-modified3}) which becomes:
\begin{equation}
	\centering
	\scalebox{0.97}[1]{$\widetilde{\text{AS}_{\text{FD}}} = P_1 \cdot \widehat{\text{AS}_{\text{DR}}} \cdot  \text{RS}_{\text{FD}} + P_2 \cdot \text{RS}_{\text{FD}} + (1-P_1) \cdot \widehat{\text{AS}_{\text{DR}}} - P_2.$}
		\label{mod1-modified3nr}
\end{equation}

Altogether, with five $\text{AS}_{\text{DR}}$/$\widehat{\text{AS}_{\text{DR}}}$ and $\text{RS}_{\text{FD}}$ combinations (Scenario 1: FSIMc-FSIMc; Scenario 2: CORNIA-FSIMc; Scenario 2: dipIQ-FSIMc; Scenario 2: NIQE-FSIMc; Scenario 2: NIQE-MSSSIM) and seven multiple distortion combinations (Blur-JPEG; Blur-Noise; JPEG-JPEG; Noise-JPEG; Noise-JP2K; NBJ-JPEG; All data), we develop 35 sets of parameter settings for Model 1.

\subsection{Model 2: Distortion Behavior Model}
\label{DRIQA_Model2}

Motivated by the simplicity of the distortion behavior analysis based 2-tier Model 1 with only two parameters, and also to better account for the non-linear behavior of certain distortion combinations such as Noise-JPEG and Noise-JP2K, we adopt a direct six-parameter polynomial model with quadratic terms as Model 2 for Scenario 1 DR IQA:
\begin{equation}
	\centering
	\begin{split}
	\widehat{\text{AS}_{\text{FD}}} & = a \cdot {\text{AS}_{\text{DR}}}^2 + b \cdot {\text{RS}_{\text{FD}}}^2 + c \cdot \text{AS}_{\text{DR}} + d \cdot \text{RS}_{\text{FD}} \\ & + e \cdot \text{AS}_{\text{DR}} \cdot \text{RS}_{\text{FD}} + f,
	\end{split}
	\label{mod2direct}
\end{equation}
where $a$, $b$, $c$, $d$, $e$ and $f$ are model coefficients, which are estimated directly using DR IQA databases V1 and V2. Model 2 reduces to Model 1 when: $a=0$, $b=0$, $c=(1-P_1)$, $d=P_2$, $e=P_1$ and $f=-P_2$. Analogous to Model 1, for the case of Scenario 2 DR IQA, Model 2 becomes:
\begin{equation}
	\centering
	\begin{split}
	\widetilde{\text{AS}_{\text{FD}}} & = a \cdot {\widehat{\text{AS}_{\text{DR}}}}^2 + b \cdot {\text{RS}_{\text{FD}}}^2 + c \cdot \widehat{\text{AS}_{\text{DR}}} + d \cdot \text{RS}_{\text{FD}} \\ & + e \cdot \widehat{\text{AS}_{\text{DR}}} \cdot \text{RS}_{\text{FD}} + f.
	\end{split}
	\label{mod2directNR}
\end{equation}
Specifically, the NR (CORNIA, dipIQ, NIQE) predicted DR image quality scores are mapped to respective FR (FSIMc or MSSSIM) scores using the nonlinear mapping function of Eq. \ref{nlmapping_ch4}. Similarly, with five $\text{AS}_{\text{DR}}$/$\widehat{\text{AS}_{\text{DR}}}$ and $\text{RS}_{\text{FD}}$ combinations, and seven multiple distortion combinations, 35 sets of parameter settings are developed for Model 2.

\subsection{Model 3: Support Vector Regression Model}
\label{DRIQA_Model3}

To better understand how well the distortion behavior based Models 1 and 2 (which use very few parameters) capture the nature of the DR IQA problem, we opt to use support vector regression (SVR) \cite{SVRbook,libsvm} to construct Model 3, which serves as an additional reference point, and also act as DR IQA models in their own right.

Specifically, we develop Model 3 by using nu-SVR that employs the radial basis function (RBF) kernel \cite{libsvm,libsvm_source,libsvm_guide} and four control parameters. For each of the 35 settings, the predictors are the FR FSIMc/MSSSIM $\text{RS}_{\text{FD}}$ scores and either the FR FSIMc $\text{AS}_{\text{DR}}$ scores or the NR CORNIA/dipIQ/NIQE $\widehat{\text{AS}_{\text{DR}}}$ scores. The training targets are the $\text{AS}_{\text{FD}}$ scores given by the SQB of the FD images. We use DR IQA database V1 for model training and DR IQA database V2 for model validation. The finalized models are later tested on a separate set of datasets (Section \ref{sec:PerfEval}). Before training, we ensure that the data has been scaled properly as recommended in \cite{libsvm_guide}. During training, we determine the best possible SVR control parameters for a particular model through an extensive grid search by training the model on DR IQA database V1 hundreds to thousands of times using different combinations of control parameters, and then selecting the parameters that lead to the best model performance in terms of both the Pearson Linear Correlation Coefficient (PLCC) and the Spearman Rank-order Correlation Coefficient (SRCC) on the validation data (DR IQA database V2). Since model training by using a large grid is quite time consuming, we use a two-tier grid search. First a coarse-level grid search is performed that identifies the region of the grid that should be focused on. This is followed by a fine-level grid search to finalize the SVR parameters which are used to train the final model on DR IQA database V1.

\section{Performance Evaluation}
\label{sec:PerfEval}
 
\subsection{Databases and Evaluation Criteria}
\label{subsec:EvalCritDB}


\begin{table*}[t!]
	\scriptsize
	\caption{Absolute performance of FR methods when determining the quality of FD images with respect to PR images in terms of PLCC.}
	\vspace{-2mm}
	\centering
	\begin{tabular}{ c | c | c | c | c | c | c | c | c }
		\hline
		\multirow{2}{*}{Database} & FR & \multicolumn{7}{c}{Distortion Combination} \\ \cline{3-9}
		 & Method & B-JPG & B-N & JPG-JPG & N-JPG & N-JP2 & NBJ-JPG & All Data \\ \hline
		 \multirowcell{2}{Waterloo Exp-II$^{\mathrm{a}}$ } & FSIMc & 0.9153 & 0.8990 & 0.9157 & 0.8932 & 0.9077 & 0.9110 & 0.9094 \\
		  & MSSSIM & 0.9363 & 0.9804 & 0.9470 & 0.8980 & 0.8989 & 0.9178 & 0.9043 \\ \hline
		 \multirowcell{2}{LIVE MD$^{\mathrm{b}}$} & FSIMc & 0.7563 & 0.7884 & -- & -- & -- & -- & 0.7690 \\
		  & MSSSIM & 0.7074 & 0.7738 & -- & -- & -- & -- & 0.6990 \\ \hline
		 \multirow{2}{*}{MDIVL$^{\mathrm{b}}$} & FSIMc & 0.8909 & -- & -- & 0.9193 & -- & -- & 0.8874 \\
		  & MSSSIM & 0.8370 & -- & -- & 0.8996 & -- & -- & 0.8645 \\ \hline
		  \multicolumn{9}{l}{$^{\mathrm{a}}$PLCC is computed with respect to SQB. \quad \quad \quad $^{\mathrm{b}}$PLCC is computed with respect to MOS/DMOS.} \\
	\end{tabular}
	\label{Table:AbsoluteFR_PLCC}
	\vspace{-2mm}
\end{table*}


\begin{table*}[t!]
	\scriptsize
	\caption{Absolute performance of FR methods when determining the quality of FD images with respect to PR images in terms of SRCC.}
	\vspace{-2mm}
	\centering
	\begin{tabular}{ c | c | c | c | c | c | c | c | c }
		\hline
		\multirow{2}{*}{Database} & FR & \multicolumn{7}{c}{Distortion Combination} \\ \cline{3-9}
		& Method & B-JPG & B-N & JPG-JPG & N-JPG & N-JP2 & NBJ-JPG & All Data \\ \hline
		\multirowcell{2}{Waterloo Exp-II$^{\mathrm{a}}$} & FSIMc & 0.9120 & 0.8956 & 0.9119 & 0.8878 & 0.9065 & 0.9076 & 0.9080 \\
		& MSSSIM & 0.9366 & 0.9809 & 0.9488 & 0.8906 & 0.9100 & 0.9204 & 0.9126 \\ \hline
		\multirowcell{2}{LIVE MD$^{\mathrm{b}}$} & FSIMc & 0.7066 & 0.7850 & -- & -- & -- & -- & 0.7517 \\
		& MSSSIM & 0.6844 & 0.7614 & -- & -- & -- & -- & 0.6941 \\ \hline
		\multirow{2}{*}{MDIVL$^{\mathrm{b}}$} & FSIMc & 0.8402 & -- & -- & 0.8765 & -- & -- & 0.8354 \\
		& MSSSIM & 0.7978 & -- & -- & 0.8175 & -- & -- & 0.8041 \\ \hline
		\multicolumn{9}{l}{$^{\mathrm{a}}$SRCC is computed with respect to SQB. \quad \quad \quad $^{\mathrm{b}}$SRCC is computed with respect to MOS/DMOS.} \\
	\end{tabular}
	\label{Table:AbsoluteFR_SRCC}
	\vspace{-2mm}
\end{table*}


\begin{table*}[htb!]
	\scriptsize
	\caption{Performance of Baseline Methods in terms of PLCC.}
	\vspace{-2mm}
	\centering
	\begin{tabular}{ c | c | c | c | c | c | c | c | c | c }
		\hline
		Baseline & \multirow{2}{*}{Database} & \multirow{2}{*}{Method} & \multicolumn{7}{c}{Distortion Combination}  \\ \cline{4-10}
		Type & & & B-JPG & B-N & JPG-JPG & N-JPG & N-JP2 & NBJ-JPG & All Data \\ \hline
		\multirowcell{8}{Baseline 1 \\ FR Methods} & \multirowcell{2}{Waterloo Exp-II$^{\mathrm{a}}$ } & FSIMc & 0.8436 & 0.8826 & 0.8276 & 0.8280 & 0.8223 & 0.7980 & 0.7926 \\
		& & MSSSIM & 0.8567 & 0.9473 & 0.8340 & 0.8809 & 0.8039 & 0.7498 & 0.7425 \\ \cline{2-10}
		& \multirowcell{2}{LIVE MD$^{\mathrm{b}}$} & FSIMc & 0.2256 & 0.3882 & -- & -- & -- & -- & 0.3045 \\
		& & MSSSIM & 0.2366 & 0.4270 & -- & -- & -- & -- & 0.2254 \\ \cline{2-10}
		& \multirow{2}{*}{MDIVL$^{\mathrm{b}}$} & FSIMc & 0.5207 & -- & -- & 0.8111 & -- & -- & 0.6238 \\
		& & MSSSIM & 0.4984 & -- & -- & 0.8770 & -- & -- & 0.5985 \\ \cline{2-10}
		& \multirowcell{2}{LIVE WCmp$^{\mathrm{b,c}}$} & FSIMc & -- & -- & -- & -- & -- & -- & 0.9030 \\					
		& & MSSSIM & -- & -- & -- & -- & -- & -- & 0.8498 \\ \hline
		\multirowcell{12}{Baseline 2 \\ NR Methods} & \multirowcell{3}{Waterloo Exp-II$^{\mathrm{a}}$} & CORNIA & 0.8918 & 0.6205 & 0.7512 & 0.7832 & 0.6943 & 0.8172 & 0.7553 \\
		& & dipIQ & 0.8522 & 0.9414 & 0.8790 & 0.8462 & 0.8380 & 0.8422 & 0.8532 \\
		& & NIQE & 0.7741 & 0.8941 & 0.7084 & 0.6368 & 0.6913 & 0.7030 & 0.7137 \\ \cline{2-10}
		& \multirowcell{3}{LIVE MD$^{\mathrm{b}}$} & CORNIA & 0.7141 & 0.8144 & -- & -- & -- & -- & 0.7360 \\
		& & dipIQ & 0.5238 & 0.6603 & -- & -- & -- & -- & 0.5531 \\
		& & NIQE & 0.7677 & 0.6670 & -- & -- & -- & -- & 0.5802 \\ \cline{2-10}
		& \multirow{3}{*}{MDIVL$^{\mathrm{b}}$} & CORNIA & 0.9331 & -- & -- & 0.7748 & -- & -- & 0.7963 \\
		& & dipIQ & 0.8298 & -- & -- & 0.8074 & -- & -- & 0.7514 \\
		& & NIQE & 0.7910 & -- & -- & 0.5357 & -- & -- & 0.5731 \\ \cline{2-10}
		& \multirowcell{3}{LIVE WCmp$^{\mathrm{b,c}}$} & CORNIA & -- & -- & -- & -- & -- & -- & 0.8424 \\
		& & dipIQ & -- & -- & -- & -- & -- & -- & 0.7978 \\
		& & NIQE & -- & -- & -- & -- & -- & -- & 0.8314 \\ \hline
		\multirowcell{4}{Baseline 3 \\ 2stepQA} & Waterloo Exp-II$^{\mathrm{a}}$ & \multirowcell{4}{2stepQA} & 0.9340 & 0.9696 & 0.8951 & 0.8420 & 0.7213 & 0.7709 & 0.7140 \\ \cline{2-2} \cline{4-10}
		& LIVE MD$^{\mathrm{b}}$ & & 0.7746 & 0.6730 & -- & -- & -- & -- & 0.6500 \\ \cline{2-2} \cline{4-10}
		& MDIVL$^{\mathrm{b}}$ & & 0.8697 & -- & -- & 0.7964 & -- & -- & 0.8149 \\ \cline{2-2} \cline{4-10}
		& LIVE WCmp$^{\mathrm{b,c}}$ & & -- & -- & -- & -- & -- & -- & 0.9229 \\ \hline
		\multicolumn{10}{l}{$^{\mathrm{a}}$PLCC is computed with respect to SQB. \quad \quad \quad $^{\mathrm{b}}$PLCC is computed with respect to MOS/DMOS.} \\
		\multicolumn{10}{l}{$^{\mathrm{c}}$LIVE WCmp has authentic distortions followed by JPEG compression. It cannot be placed in a specific distortion combination.} \\
	\end{tabular}
	\label{Table:Baseline_PLCC}
	\vspace{-2mm}
\end{table*}


\begin{table*}[htb!]
	\scriptsize
	\caption{Performance of Baseline Methods in terms of SRCC.}
	\vspace{-2mm}
	\centering
	\begin{tabular}{ c | c | c | c | c | c | c | c | c | c }
		\hline
		Baseline & \multirow{2}{*}{Database} & \multirow{2}{*}{Method} & \multicolumn{7}{c}{Distortion Combination}  \\ \cline{4-10}
		Type & & & B-JPG & B-N & JPG-JPG & N-JPG & N-JP2 & NBJ-JPG & All Data \\ \hline
		\multirowcell{8}{Baseline 1 \\ FR Methods} & \multirowcell{2}{Waterloo Exp-II$^{\mathrm{a}}$ } & FSIMc & 0.8442 & 0.8843 & 0.7544 & 0.8195 & 0.8154 & 0.7964 & 0.7816 \\
		& & MSSSIM & 0.8561 & 0.9481 & 0.7467 & 0.8803 & 0.7972 & 0.7579 & 0.7293 \\ \cline{2-10}
		& \multirowcell{2}{LIVE MD$^{\mathrm{b}}$} & FSIMc & 0.1923 & 0.3336 & -- & -- & -- & -- & 0.2446 \\
		& & MSSSIM & 0.1370 & 0.3671 & -- & -- & -- & -- & 0.2076 \\ \cline{2-10}
		& \multirow{2}{*}{MDIVL$^{\mathrm{b}}$} & FSIMc & 0.4870 & -- & -- & 0.8243 & -- & -- & 0.6316 \\
		& & MSSSIM & 0.4619 & -- & -- & 0.8781 & -- & -- & 0.5698 \\ \cline{2-10}
		& \multirowcell{2}{LIVE WCmp$^{\mathrm{b,c}}$} & FSIMc & -- & -- & -- & -- & -- & -- & 0.9024 \\						
		& & MSSSIM & -- & -- & -- & -- & -- & -- & 0.8469 \\ \hline
		\multirowcell{12}{Baseline 2 \\ NR Methods} & \multirowcell{3}{Waterloo Exp-II$^{\mathrm{a}}$} & CORNIA & 0.8955 & 0.5940 & 0.7539 & 0.7826 & 0.7131 & 0.8101 & 0.7576 \\
		& & dipIQ & 0.8508 & 0.9443 & 0.8786 & 0.8403 & 0.8549 & 0.8437 & 0.8516 \\
		& & NIQE & 0.7670 & 0.8939 & 0.6978 & 0.6150 & 0.6975 & 0.6890 & 0.6891 \\ \cline{2-10}
		& \multirowcell{3}{LIVE MD$^{\mathrm{b}}$} & CORNIA & 0.6897 & 0.7997 & -- & -- & -- & -- & 0.7278 \\
		& & dipIQ & 0.4823 & 0.5706 & -- & -- & -- & -- & 0.4548 \\
		& & NIQE & 0.7487 & 0.6359 & -- & -- & -- & -- & 0.5512 \\ \cline{2-10}
		& \multirow{3}{*}{MDIVL$^{\mathrm{b}}$} & CORNIA & 0.9202 & -- & -- & 0.8101 & -- & -- & 0.8157 \\
		& & dipIQ & 0.6561 & -- & -- & 0.8393 & -- & -- & 0.7423 \\
		& & NIQE & 0.7558 & -- & -- & 0.5791 & -- & -- & 0.5946 \\ \cline{2-10}
		& \multirowcell{3}{LIVE WCmp$^{\mathrm{b,c}}$} & CORNIA & -- & -- & -- & -- & -- & -- & 0.8471 \\
		& & dipIQ & -- & -- & -- & -- & -- & -- & 0.7868 \\
		& & NIQE & -- & -- & -- & -- & -- & -- & 0.8327 \\ \hline
		\multirowcell{4}{Baseline 3 \\ 2stepQA} & Waterloo Exp-II$^{\mathrm{a}}$ & \multirowcell{4}{2stepQA} & 0.9337 & 0.9708 & 0.8669 & 0.8342 & 0.7162 & 0.7797 & 0.7274 \\ \cline{2-2} \cline{4-10}
		& LIVE MD$^{\mathrm{b}}$ &  & 0.7530 & 0.5356 & -- & -- & -- & -- & 0.5318 \\ \cline{2-2} \cline{4-10}
		& MDIVL$^{\mathrm{b}}$ &  & 0.8539 & -- & -- & 0.7685 & -- & -- & 0.7713 \\ \cline{2-2} \cline{4-10}
		& LIVE WCmp$^{\mathrm{b,c}}$ &  & -- & -- & -- & -- & -- & -- & 0.9246 \\ \hline
		\multicolumn{10}{l}{$^{\mathrm{a}}$SRCC is computed with respect to SQB. \quad \quad \quad $^{\mathrm{b}}$SRCC is computed with respect to MOS/DMOS.} \\
		\multicolumn{10}{l}{$^{\mathrm{c}}$LIVE WCmp has authentic distortions followed by JPEG compression. It cannot be placed in a specific distortion combination.} \\
	\end{tabular}
	\label{Table:Baseline_SRCC}
	\vspace{-2mm}
\end{table*}

To the best of our knowledge, there are only four datasets \cite{dnn_database_tip,md_2stepQA_jrnl,db_livemd,db_mdivl} that provide both singly distorted DR and their respective multiply distorted FD images, together with quality labels. Although two other datasets, MDID \cite{db_mdid} and MDID2013 \cite{md_sisblim_db_mdid2013}, contain multiply distorted images, they do not provide DR images. Therefore, we use these four datasets, as discussed below, for performance evaluation. These datasets do not have any content overlap with DR IQA databases V1 and V2 used in the development of our DR IQA models.

The Waterloo Exploration-II (Waterloo Exp-II) database \cite{dnn_database_tip} has 3,570 PR images, 39,270 singly distorted images each for Blur, JPEG compression, and Noise, and 667,590 multiply distorted images each for the distortion combinations of Blur-JPEG, Blur-Noise, JPEG-JPEG, Noise-JPEG, and Noise-JP2K. The singly and multiply distorted images are the DR and FD images, respectively, in a 2-stage distortion process. All distorted images are annotated with synthetic quality benchmark (SQB) labels that have been generated by fusing the results from four state-of-the-art FR methods~\cite{dnn_database_tip}.

The LIVE Wild Compressed (LIVE WCmp) database \cite{md_2stepQA_conf,md_2stepQA_jrnl,db_liveWComp_source} is composed of 400 images. It starts with 80 authentically distorted images taken from the LIVE Wild Challenge database \cite{db_livewc} which can be regarded as DR images. Each of these 80 images are further JPEG compressed at four fixed compression levels regardless of content, leading to 320 FD images. LIVE WCmp provides subjective ratings for all its images in the form of mean opinion scores (MOS). The dataset does not have PR images.

The LIVE Multiply Distorted (LIVE MD) database \cite{db_livemd,db_livemd_source} consists of 15 PR images, 45 singly distorted images each for Blur, JPEG compression and Noise, and 135 multiply distorted images each for the distortion combinations of Blur-JPEG and Blur-Noise. We consider the singly distorted Blur images as DR and the multiply distorted images as the FD images. Subjective ratings are available in the form of difference mean opinion scores (DMOS).

The Multiply Distorted IVL (MDIVL) database \cite{db_mdivl,db_mdivl_copyrightNotice,db_mdivl_source} consists of 10 PR and 750 multiply distorted images of which 350 belong to the Blur-JPEG combination while 400 belong to the Noise-JPEG combination. Although the database does not explicitly contain singly distorted images, in both Blur-JPEG and Noise-JPEG combinations, the least compression distortion level utilizes MATLAB quality factor of 100, which produces nearly perceptually lossless compression. Thus, we regard 70 out of 350 Blur-JPEG and 100 out of 400 Noise-JPEG images as singly distorted Blur and Noise images, respectively, thereby providing us with DR and FD images. MDIVL provides subjective ratings for all of its distorted images in the form of MOS.
 
We use PLCC and SRCC as measures of a model's prediction accuracy and prediction monotonicity, respectively \cite{vqeg_report}. PLCC is computed after a nonlinear mapping step between model predictions and target scores, whereas SRCC is computed directly \cite{eval_Waterloo}.

\subsection{Absolute FR Performance and Baseline IQA Models}
\label{subsec:BasePerf}

With access to the pristine reference images, FR IQA methods offer the best quality prediction performance~\cite{eval_Waterloo}, and thus serve as an approximate upper bound for the baseline and DR IQA models that we will test later. Here we select FSIMc \cite{fr_fsim} and MSSSIM \cite{fr_msssim} as the reference FR models. The former outperforms most other FR methods \cite{eval_Waterloo}, and the latter is the FR component in 2stepQA \cite{md_2stepQA_jrnl} besides being a competitive method \cite{eval_Waterloo}. Tables \ref{Table:AbsoluteFR_PLCC} and \ref{Table:AbsoluteFR_SRCC} show the performance of the FR methods in terms of PLCC and SRCC, respectively, where the results are termed as absolute performance since testing is done against PR images. The LIVE WCmp database \cite{md_2stepQA_jrnl} is not present because it does not have PR images.

Given the two-stage DR IQA framework, three types of approaches may be applied using existing IQA models in the literature, and are included as baseline models. 1) Baseline-1: Use FR methods to assess the relative quality of FD images with respect to DR images, i.e., $\text{RS}_{\text{FD}}$, and use it to predict the absolute quality of FD images. Specifically, we use FSIMc \cite{fr_fsim} and MSSSIM \cite{fr_msssim} as the representative FR measures; 2) Baseline-2: Use NR methods to assess the FD images directly, without referencing to the DR images. In particular, CORNIA \cite{nr_cornia} and dipIQ \cite{nr_dipiq} are selected as they were found to be the top performers in \cite{eval_Waterloo}, while NIQE \cite{nr_niqe} is selected as it is the NR component in 2stepQA \cite{md_2stepQA_jrnl}; 3) Baseline-3: 2stepQA \cite{md_2stepQA_conf,md_2stepQA_jrnl} is the only model that utilizes the quality information about the DR image while determining the quality of a multiply distorted FD image. Specifically, NIQE \cite{nr_niqe} is used for NR assessment of the DR image, MSSSIM \cite{fr_msssim} is used to compare the DR and FD images, and a product of the scores produces a final quality assessment of the FD image.

Tables \ref{Table:Baseline_PLCC} and \ref{Table:Baseline_SRCC} provide the quality prediction performance of the three baseline approaches in terms of PLCC and SRCC, respectively. There are several important observations. First, close comparison with Tables \ref{Table:AbsoluteFR_PLCC} and \ref{Table:AbsoluteFR_SRCC} shows that there is generally a large gap in performance between the absolute FR quality scores and the Baseline-1 relative FR quality scores, even though the same top-performing FR models are employed in both cases, suggesting that FR IQA models are reliable only when the pristine quality reference is accessible. Second, the Baseline-2 NR models, including the state-of-the-art CORNIA \cite{nr_cornia}, perform highly inconsistently and largely depend on the test dataset, the distortion types, and the NR model being used. Since most NR models are developed, trained and/or validated with singly distorted images, there is generally a large performance drop when they evaluate multiply distorted images. Furthermore, as shown in \cite{eval_Waterloo}, NR methods SISBLIM \cite{md_sisblim_db_mdid2013} and GWHGLBP \cite{md_gwhglbp}, that are designed for multiply distorted content, are unable to outperform CORNIA \cite{nr_cornia} even on multiply distorted datasets, and their performance drops further when testing also incorporates singly distorted content. Such inconsistent performance of the NR IQA paradigm, owing to the difficult nature of the problem and its inability to use auxiliary information about a distorted image even if it is available, is a strong motivation for the development of the new DR IQA paradigm that is able to utilize additional information provided by DR images. Third, the Baseline-3 2stepQA model may sometimes significantly improve upon the first 2 Baseline models, but the performance gain varies drastically, and is mostly limited to certain types of distortion combinations such as B-JPG and JPG-JPG. Overall, all three baseline models exhibit significant performance gaps against the absolute FR IQA scores shown in Tables \ref{Table:AbsoluteFR_PLCC} and \ref{Table:AbsoluteFR_SRCC}, suggesting the potential space for improvement by deeper investigation on DR IQA.

\subsection{Performance of DR IQA Models}
\label{subsec:DRIQA_Perf}


\begin{table*}[t!]
	\scriptsize
	\caption{Performance of DR IQA Models in terms of PLCC.}
	\vspace{-2mm}
	\centering
		\begin{tabular}{ c | c | c | c | c | c | c | c | c | c | c }
			\hline
			DR IQA & \multirow{2}{*}{Database} & \multicolumn{2}{c|}{Predictors} & \multicolumn{7}{c}{Distortion Combination and Model Type}  \\ \cline{3-11}
			Model & & $\text{AS}_{\text{DR}}$/$\widehat{\text{AS}_{\text{DR}}}$ & $\text{RS}_{\text{FD}}$ & B-JPG & B-N & JPG-JPG & N-JPG & N-JP2 & NBJ-JPG & All Data \\ \hline
			\multirowcell{19}{Model 1 \\ (Distortion \\ Behavior \\ Based)} & \multirowcell{5}{Waterloo Exp-II$^{\mathrm{a}}$} & FSIMc & FSIMc & 0.9126 & 0.8991 & 0.9201 & 0.8568 & 0.8147 & 0.8538 & 0.8217 \\ \cline{3-11}
			&  & CORNIA & FSIMc & 0.9085 & 0.9119 & 0.9088 & 0.8559 & 0.8291 & 0.8540 & 0.8264 \\
			&  & dipIQ & FSIMc & 0.9079 & 0.9114 & 0.9219 & 0.8922 & 0.7998 & 0.8653 & 0.8335 \\
			&  & NIQE & FSIMc & 0.8917 & 0.9041 & 0.8860 & 0.8626 & 0.8041 & 0.8367 & 0.8142 \\
			&  & NIQE & MSSSIM & 0.9341 & 0.9682 & 0.9165 & 0.7815 & 0.5839 & 0.7754 & 0.7628 \\ \cline{2-11}
			& \multirowcell{5}{LIVE MD$^{\mathrm{b}}$} & FSIMc & FSIMc & 0.7576 & 0.7911 & -- & -- & -- & -- & 0.7662 \\ \cline{3-11}
			&  & CORNIA & FSIMc & 0.7786 & 0.7616 & -- & -- & -- & -- & 0.7635 \\
			&  & dipIQ & FSIMc & 0.7797 & 0.7932 & -- & -- & -- & -- & 0.7797 \\
			&  & NIQE & FSIMc & 0.7827 & 0.7677 & -- & -- & -- & -- & 0.7637 \\
			&  & NIQE & MSSSIM & 0.7744 & 0.6818 & -- & -- & -- & -- & 0.6705 \\ \cline{2-11}			
			& \multirow{5}{*}{MDIVL$^{\mathrm{b,c}}$} & FSIMc & FSIMc & 0.8964 & -- & -- & 0.9179 & -- & 0.9001 & 0.8997 \\ \cline{3-11}
			&  & CORNIA & FSIMc & 0.9192 & -- & -- & 0.8551 & -- & 0.8858 & 0.8859 \\
			&  & dipIQ & FSIMc & 0.9167 & -- & -- & 0.8928 & -- & 0.8921 & 0.8921 \\
			&  & NIQE & FSIMc & 0.8605 & -- & -- & 0.7856 & -- & 0.8234 & 0.8223 \\
			&  & NIQE & MSSSIM & 0.8607 & -- & -- & 0.7430 & -- & 0.8091 & 0.8075 \\ \cline{2-11}
			& \multirowcell{4}{LIVE WCmp$^{\mathrm{b,d}}$} & CORNIA & FSIMc & 0.9100 & 0.9096 & 0.9084 & 0.9055 & 0.8971 & 0.9095 & 0.9094 \\
			&  & dipIQ & FSIMc & 0.9081 & 0.9080 & 0.9080 & 0.9076 & 0.9071 & 0.9080 & 0.9079 \\
			&  & NIQE & FSIMc & 0.9280 & 0.9271 & 0.9271 & 0.9218 & 0.9139 & 0.9271 & 0.9258 \\
			&  & NIQE & MSSSIM & 0.9264 & 0.9261 & 0.9259 & 0.9175 & 0.9134 & 0.9156 & 0.9133 \\ \hline
			\multirowcell{19}{Model 2 \\ (Distortion \\ Behavior \\ Based)} & \multirowcell{5}{Waterloo Exp-II$^{\mathrm{a}}$} & FSIMc & FSIMc & 0.9135 & 0.9003 & 0.9206 & 0.8751 & 0.8857 & 0.8567 & 0.8296 \\ \cline{3-11}
			&  & CORNIA & FSIMc & 0.9090 & 0.9117 & 0.9085 & 0.8654 & 0.8432 & 0.8550 & 0.8288 \\
			&  & dipIQ & FSIMc & 0.9078 & 0.9116 & 0.9228 & 0.9075 & 0.8751 & 0.8685 & 0.8416 \\
			&  & NIQE & FSIMc & 0.8911 & 0.9042 & 0.8854 & 0.8723 & 0.8606 & 0.8414 & 0.8248 \\
			&  & NIQE & MSSSIM & 0.9336 & 0.9686 & 0.9182 & 0.8726 & 0.8490 & 0.8132 & 0.7980 \\ \cline{2-11}
			& \multirowcell{5}{LIVE MD$^{\mathrm{b}}$} & FSIMc & FSIMc & 0.7575 & 0.7911 & -- & -- & -- & -- & 0.7628 \\ \cline{3-11}
			&  & CORNIA & FSIMc & 0.7745 & 0.7679 & -- & -- & -- & -- & 0.7901 \\
			&  & dipIQ & FSIMc & 0.7797 & 0.7963 & -- & -- & -- & -- & 0.7772 \\
			&  & NIQE & FSIMc & 0.7825 & 0.7736 & -- & -- & -- & -- & 0.7615 \\
			&  & NIQE & MSSSIM & 0.7071 & 0.7089 & -- & -- & -- & -- & 0.6347 \\ \cline{2-11}
			& \multirow{5}{*}{MDIVL$^{\mathrm{b,c}}$} & FSIMc & FSIMc & 0.8970 & -- & -- & 0.9221 & -- & 0.9008 & 0.8990 \\ \cline{3-11}
			&  & CORNIA & FSIMc & 0.9147 & -- & -- & 0.8793 & -- & 0.8953 & 0.8958 \\
			&  & dipIQ & FSIMc & 0.9164 & -- & -- & 0.9076 & -- & 0.8906 & 0.8863 \\
			&  & NIQE & FSIMc & 0.8586 & -- & -- & 0.7988 & -- & 0.8202 & 0.8167 \\
			&  & NIQE & MSSSIM & 0.8572 & -- & -- & 0.7633 & -- & 0.7874 & 0.7550 \\ \cline{2-11}
			& \multirowcell{4}{LIVE WCmp$^{\mathrm{b,d}}$} & CORNIA & FSIMc & 0.9084 & 0.9109 & 0.9093 & 0.9122 & 0.9035 & 0.9141 & 0.9140 \\
			&  & dipIQ & FSIMc & 0.9081 & 0.9079 & 0.9081 & 0.9058 & 0.9058 & 0.9079 & 0.9077 \\
			&  & NIQE & FSIMc & 0.9277 & 0.9271 & 0.9265 & 0.9201 & 0.9100 & 0.9255 & 0.9233 \\
			&  & NIQE & MSSSIM & 0.9255 & 0.9258 & 0.9242 & 0.9067 & 0.8886 & 0.9187 & 0.9134 \\ \hline
			\multirowcell{19}{Model 3 \\ (SVR \\ Based)} & \multirowcell{5}{Waterloo Exp-II$^{\mathrm{a}}$} & FSIMc & FSIMc & 0.9287 & 0.9104 & 0.9195 & 0.8877 & 0.9074 & 0.8629 & 0.8416 \\ \cline{3-11}
			&  & CORNIA & FSIMc & 0.9215 & 0.9180 & 0.9062 & 0.8547 & 0.8449 & 0.8643 & 0.8389 \\
			&  & dipIQ & FSIMc & 0.9228 & 0.9195 & 0.9224 & 0.9112 & 0.8825 & 0.8690 & 0.8448 \\
			&  & NIQE & FSIMc & 0.9017 & 0.9089 & 0.8853 & 0.8809 & 0.8660 & 0.8422 & 0.8317 \\
			&  & NIQE & MSSSIM & 0.9383 & 0.9671 & 0.9159 & 0.9327 & 0.8746 & 0.8172 & 0.7952 \\ \cline{2-11}
			& \multirowcell{5}{LIVE MD$^{\mathrm{b}}$} & FSIM & FSIMc & 0.7539 & 0.8082 & -- & -- & -- & -- & 0.7329 \\ \cline{3-11}
			&  & CORNIA & FSIMc & 0.7769 & 0.8175 & -- & -- & -- & -- & 0.7032 \\
			&  & dipIQ & FSIMc & 0.7371 & 0.7791 & -- & -- & -- & -- & 0.7839 \\
			&  & NIQE & FSIMc & 0.7712 & 0.7641 & -- & -- & -- & -- & 0.7602 \\
			&  & NIQE & MSSSIM & 0.7349 & 0.7270 & -- & -- & -- & -- & 0.6468 \\ \cline{2-11}
			& \multirow{5}{*}{MDIVL$^{\mathrm{b,c}}$} & FSIMc & FSIMc & 0.8975 & -- & -- & 0.9227 & -- & 0.9063 & 0.9048 \\ \cline{3-11}
			&  & CORNIA & FSIMc & 0.9397 & -- & -- & 0.8767 & -- & 0.9085 & 0.9001 \\
			&  & dipIQ & FSIMc & 0.9203 & -- & -- & 0.9052 & -- & 0.8999 & 0.8950 \\
			&  & NIQE & FSIMc & 0.8578 & -- & -- & 0.8182 & -- & 0.8296 & 0.8046 \\
			&  & NIQE & MSSSIM & 0.8671 & -- & -- & 0.8161 & -- & 0.7737 & 0.7799 \\ \cline{2-11}
			& \multirowcell{4}{LIVE WCmp$^{\mathrm{b,d}}$} & CORNIA & FSIMc & 0.9117 & 0.9156 & 0.9166 & 0.9087 & 0.9023 & 0.9166 & 0.9133 \\
			&  & dipIQ & FSIMc & 0.9086 & 0.9083 & 0.9065 & 0.9010 & 0.9010 & 0.9067 & 0.9069 \\
			&  & NIQE & FSIMc & 0.9239 & 0.9237 & 0.9207 & 0.9169 & 0.9056 & 0.9176 & 0.9202 \\
			&  & NIQE & MSSSIM & 0.9247 & 0.9217 & 0.9188 & 0.8609 & 0.8514 & 0.9152 & 0.9131 \\ \hline
			\multicolumn{11}{l}{$^{\mathrm{a}}$PLCC is computed with respect to SQB. \quad \quad \quad $^{\mathrm{b}}$PLCC is computed with respect to MOS/DMOS.} \\
			\multicolumn{11}{l}{$^{\mathrm{c}}$ The NBJ-JPG and All Data model versions are applied to the entire MDIVL database.} \\
			\multicolumn{11}{l}{$^{\mathrm{d}}$LIVE WCmp has authentic distortions followed by JPEG compression. It cannot be placed in a specific distortion combination. Models,} \\
			\multicolumn{11}{l}{\hspace{1mm}   trained for the seven distortion combinations, are applied to the entire dataset and results have been reported in respective columns.} \\
		\end{tabular}
	\label{Table:DR_IQA_Models_PLCC}
	\vspace{-2mm}
\end{table*}


\begin{table*}[t!]
	\scriptsize
	\caption{Performance of DR IQA Models in terms of SRCC.}
	\vspace{-2mm}
	\centering
		\begin{tabular}{ c | c | c | c | c | c | c | c | c | c | c }
			\hline
			DR IQA & \multirow{2}{*}{Database} & \multicolumn{2}{c|}{Predictors} & \multicolumn{7}{c}{Distortion Combination and Model Type}  \\ \cline{3-11}
			Model & & $\text{AS}_{\text{DR}}$/$\widehat{\text{AS}_{\text{DR}}}$ & $\text{RS}_{\text{FD}}$ & B-JPG & B-N & JPG-JPG & N-JPG & N-JP2 & NBJ-JPG & All Data \\ \hline
			\multirowcell{19}{Model 1 \\ (Distortion \\ Behavior \\ Based)} & \multirowcell{5}{Waterloo Exp-II$^{\mathrm{a}}$} & FSIMc & FSIMc & 0.9094 & 0.8964 & 0.9145 & 0.8541 & 0.8032 & 0.8653 & 0.8209 \\ \cline{3-11}
			&  & CORNIA & FSIMc & 0.9067 & 0.9117 & 0.8905 & 0.8512 & 0.8223 & 0.8579 & 0.8204 \\
			&  & dipIQ & FSIMc & 0.9069 & 0.9111 & 0.9107 & 0.8940 & 0.8071 & 0.8764 & 0.8324 \\
			&  & NIQE & FSIMc & 0.8905 & 0.9045 & 0.8597 & 0.8489 & 0.7987 & 0.8433 & 0.8080 \\
			&  & NIQE & MSSSIM & 0.9348 & 0.9690 & 0.8831 & 0.7568 & 0.5181 & 0.7804 & 0.7490 \\ \cline{2-11}			
			& \multirowcell{5}{LIVE MD$^{\mathrm{b}}$} & FSIMc & FSIMc & 0.7067 & 0.7876 & -- & -- & -- & -- & 0.7470 \\ \cline{3-11}
			&  & CORNIA & FSIMc & 0.7589 & 0.7021 & -- & -- & -- & -- & 0.7306 \\
			&  & dipIQ & FSIMc & 0.7292 & 0.7576 & -- & -- & -- & -- & 0.7372 \\
			&  & NIQE & FSIMc & 0.7641 & 0.7241 & -- & -- & -- & -- & 0.7396 \\
			&  & NIQE & MSSSIM & 0.7518 & 0.5536 & -- & -- & -- & -- & 0.5685 \\ \cline{2-11}
			& \multirow{5}{*}{MDIVL$^{\mathrm{b,c}}$} & FSIMc & FSIMc & 0.8436 & -- & -- & 0.8741 & -- & 0.8493 & 0.8466 \\ \cline{3-11}
			&  & CORNIA & FSIMc & 0.8941 & -- & -- & 0.8505 & -- & 0.8749 & 0.8772 \\
			&  & dipIQ & FSIMc & 0.8947 & -- & -- & 0.8435 & -- & 0.8628 & 0.8621 \\
			&  & NIQE & FSIMc & 0.8408 & -- & -- & 0.7542 & -- & 0.8030 & 0.8024 \\
			&  & NIQE & MSSSIM & 0.8450 & -- & -- & 0.7005 & -- & 0.7646 & 0.7617 \\ \cline{2-11}
			& \multirowcell{4}{LIVE WCmp$^{\mathrm{b,d}}$} & CORNIA & FSIMc & 0.9152 & 0.9134 & 0.9142 & 0.9069 & 0.8938 & 0.9136 & 0.9124 \\
			&  & dipIQ & FSIMc & 0.9097 & 0.9094 & 0.9096 & 0.9083 & 0.9077 & 0.9094 & 0.9091 \\
			&  & NIQE & FSIMc & 0.9287 & 0.9266 & 0.9274 & 0.9189 & 0.9091 & 0.9264 & 0.9245 \\
			&  & NIQE & MSSSIM & 0.9284 & 0.9282 & 0.9282 & 0.9163 & 0.9110 & 0.9158 & 0.9129 \\ \hline
			\multirowcell{19}{Model 2 \\ (Distortion \\ Behavior \\ Based)} & \multirowcell{5}{Waterloo Exp-II$^{\mathrm{a}}$} & FSIMc & FSIMc & 0.9104 & 0.8975 & 0.9157 & 0.8701 & 0.8685 & 0.8680 & 0.8279 \\ \cline{3-11}
			&  & CORNIA & FSIMc & 0.9075 & 0.9115 & 0.8897 & 0.8581 & 0.8369 & 0.8580 & 0.8225 \\
			&  & dipIQ & FSIMc & 0.9067 & 0.9112 & 0.9116 & 0.9077 & 0.8737 & 0.8791 & 0.8393 \\
			&  & NIQE & FSIMc & 0.8898 & 0.9046 & 0.8563 & 0.8613 & 0.8525 & 0.8468 & 0.8173 \\
			&  & NIQE & MSSSIM & 0.9343 & 0.9693 & 0.8837 & 0.8432 & 0.8357 & 0.8152 & 0.7850 \\ \cline{2-11}
			& \multirowcell{5}{LIVE MD$^{\mathrm{b}}$} & FSIMc & FSIMc & 0.7069 & 0.7865 & -- & -- & -- & -- & 0.7376 \\ \cline{3-11}
			&  & CORNIA & FSIMc & 0.7546 & 0.7162 & -- & -- & -- & -- & 0.7795 \\
			&  & dipIQ & FSIMc & 0.7294 & 0.7625 & -- & -- & -- & -- & 0.7351 \\
			&  & NIQE & FSIMc & 0.7646 & 0.7338 & -- & -- & -- & -- & 0.7378 \\
			&  & NIQE & MSSSIM & 0.6761 & 0.6176 & -- & -- & -- & -- & 0.387 \\ \cline{2-11}
			& \multirow{5}{*}{MDIVL$^{\mathrm{b,c}}$} & FSIMc & FSIMc & 0.8464 & -- & -- & 0.8889 & -- & 0.8486 & 0.8426 \\ \cline{3-11}
			&  & CORNIA & FSIMc & 0.8894 & -- & -- & 0.8882 & -- & 0.8873 & 0.8889 \\
			&  & dipIQ & FSIMc & 0.8946 & -- & -- & 0.8691 & -- & 0.8642 & 0.8613 \\
			&  & NIQE & FSIMc & 0.8396 & -- & -- & 0.7802 & -- & 0.8035 & 0.8023 \\
			&  & NIQE & MSSSIM & 0.8413 & -- & -- & 0.7143 & -- & 0.7595 & 0.7239 \\ \cline{2-11}
			& \multirowcell{4}{LIVE WCmp$^{\mathrm{b,d}}$} & CORNIA & FSIMc & 0.9139 & 0.9145 & 0.9143 & 0.9121 & 0.9026 & 0.9168 & 0.9154 \\
			&  & dipIQ & FSIMc & 0.9097 & 0.9093 & 0.9095 & 0.9055 & 0.9056 & 0.9092 & 0.9087 \\
			&  & NIQE & FSIMc & 0.9286 & 0.9264 & 0.9266 & 0.9182 & 0.9063 & 0.9247 & 0.9214 \\
			&  & NIQE & MSSSIM & 0.9275 & 0.9279 & 0.9258 & 0.9052 & 0.8880 & 0.9190 & 0.9148 \\ \hline
			\multirowcell{19}{Model 3 \\ (SVR \\ Based)} & \multirowcell{5}{Waterloo Exp-II$^{\mathrm{a}}$} & FSIMc & FSIMc & 0.9286 & 0.9079 & 0.9155 & 0.8794 & 0.8957 & 0.8753 & 0.8389 \\ \cline{3-11}
			&  & CORNIA & FSIMc & 0.9228 & 0.9181 & 0.8861 & 0.8511 & 0.8386 & 0.8734 & 0.8391 \\
			&  & dipIQ & FSIMc & 0.9252 & 0.9194 & 0.9109 & 0.9118 & 0.8809 & 0.8800 & 0.8434 \\
			&  & NIQE & FSIMc & 0.9038 & 0.9095 & 0.8558 & 0.8712 & 0.8601 & 0.8474 & 0.8230 \\
			&  & NIQE & MSSSIM & 0.9387 & 0.9689 & 0.8815 & 0.9276 & 0.8670 & 0.8255 & 0.7852 \\ \cline{2-11}
			& \multirowcell{5}{LIVE MD$^{\mathrm{b}}$} & FSIMc & FSIMc & 0.7154 & 0.7983 & -- & -- & -- & -- & 0.7015 \\ \cline{3-11}
			&  & CORNIA & FSIMc & 0.7613 & 0.8022 & -- & -- & -- & -- & 0.6533 \\
			&  & dipIQ & FSIMc & 0.6890 & 0.7105 & -- & -- & -- & -- & 0.7545 \\
			&  & NIQE & FSIMc & 0.7478 & 0.7229 & -- & -- & -- & -- & 0.7427 \\
			&  & NIQE & MSSSIM & 0.7110 & 0.6575 & -- & -- & -- & -- & 0.4111 \\ \cline{2-11}
			& \multirow{5}{*}{MDIVL$^{\mathrm{b,c}}$} & FSIMc & FSIMc & 0.8266 & -- & -- & 0.9020 & -- & 0.8758 & 0.8671 \\ \cline{3-11}
			&  & CORNIA & FSIMc & 0.9145 & -- & -- & 0.8871 & -- & 0.9020 & 0.8910 \\
			&  & dipIQ & FSIMc & 0.8731 & -- & -- & 0.8776 & -- & 0.8759 & 0.8726 \\
			&  & NIQE & FSIMc & 0.8295 & -- & -- & 0.8048 & -- & 0.8211 & 0.7872 \\
			&  & NIQE & MSSSIM & 0.8537 & -- & -- & 0.8092 & -- & 0.7239 & 0.7247 \\ \cline{2-11}
			& \multirowcell{4}{LIVE WCmp$^{\mathrm{b,d}}$} & CORNIA & FSIMc & 0.9150 & 0.9182 & 0.9184 & 0.9087 & 0.9000 & 0.9156 & 0.9145 \\
			&  & dipIQ & FSIMc & 0.9104 & 0.9098 & 0.9059 & 0.8939 & 0.8941 & 0.9061 & 0.9045 \\
			&  & NIQE & FSIMc & 0.9220 & 0.9219 & 0.9159 & 0.9155 & 0.9005 & 0.9143 & 0.9184 \\
			&  & NIQE & MSSSIM & 0.9246 & 0.9235 & 0.9158 & 0.8622 & 0.8546 & 0.9126 & 0.9120 \\ \hline
			\multicolumn{11}{l}{$^{\mathrm{a}}$SRCC is computed with respect to SQB. \quad \quad \quad $^{\mathrm{b}}$SRCC is computed with respect to MOS/DMOS.} \\
			\multicolumn{11}{l}{$^{\mathrm{c}}$ The NBJ-JPG and All Data model versions are applied to the entire MDIVL database.} \\
			\multicolumn{11}{l}{$^{\mathrm{d}}$LIVE WCmp has authentic distortions followed by JPEG compression. It cannot be placed in a specific distortion combination. Models,} \\
			\multicolumn{11}{l}{\hspace{1mm}   trained for the seven distortion combinations, are applied to the entire dataset and results have been reported in respective columns.} \\
			\multicolumn{11}{l}{\hspace{1mm}  } \\
		\end{tabular}
	\label{Table:DR_IQA_Models_SRCC}
	\vspace{-2mm}
\end{table*}

Tables \ref{Table:DR_IQA_Models_PLCC} and \ref{Table:DR_IQA_Models_SRCC} provide the performance of DR IQA Models 1, 2, and 3 in terms of PLCC and SRCC, respectively, where the test datasets have no content overlap with DR IQA databases V1 and V2 used for model development.

\subsubsection{Comparison with Baseline Models}
\label{DRIQA_PerfWRTbaseline}

A comparison of Tables \ref{Table:DR_IQA_Models_PLCC} and \ref{Table:DR_IQA_Models_SRCC} with Tables \ref{Table:Baseline_PLCC} and \ref{Table:Baseline_SRCC}, respectively, shows that DR IQA Models 1, 2, and 3 outperform the FR based Baseline-1 approach significantly and nearly comprehensively. Their superior performance relative to Baseline-1 demonstrates the shortcomings of the FR paradigm in the absence of PR images at the final destination and establishes the value of the DR IQA framework.

Comparing Tables \ref{Table:DR_IQA_Models_PLCC} and \ref{Table:DR_IQA_Models_SRCC} with Tables \ref{Table:AbsoluteFR_PLCC} and \ref{Table:AbsoluteFR_SRCC}, respectively, we find that the DR IQA models perform better than or at par in most cases against FR computed $\text{AS}_{\text{FD}}$ scores. This is no small achievement given that FR performance is usually considered as an upper bound in IQA when the PR images are accessible. There are a few exceptions on the N-JP2, NBJ-JPG and All data cases. This highlights the difficult nature of the N-JP2 case, as can be seen in the distortion behavior plot of Fig. \ref{fig:Plots_ASfd_vs_RSfd}(e). It also highlights the difficult nature of the NBJ-JPG and All data cases, where multiple distortion combinations are considered together.

Comparing Tables \ref{Table:DR_IQA_Models_PLCC} and \ref{Table:DR_IQA_Models_SRCC} with Tables \ref{Table:Baseline_PLCC} and \ref{Table:Baseline_SRCC}, respectively, shows that DR IQA Models nearly comprehensively outperform the NR based Baseline-2 approach on all test datasets. A few NR models perform exceptionally well on certain test cases (e.g., CORNIA \cite{nr_cornia} on B-N and B-JPG cases of LIVE MD and MDIVL, respectively, and dipIQ \cite{nr_dipiq} on Waterloo Exp-II), but their performance drops drastically on other cases. These results suggest the benefit of incorporating the additional information in the DR images, and again demonstrate the value of the DR IQA paradigm. The superior performance of Scenario 2 or Type-001100 DR IQA, compared to Baseline-2, also shows that in the absence of PR images, NR methods can be effectively used to compute $\widehat{\text{AS}_{\text{DR}}}$ scores for DR images, which together with the FR computed $\text{RS}_{\text{FD}}$ scores between the DR and FD images, can lead to effective DR IQA models, again highlighting the value of using additional information provided by DR images even if it is through their NR predicted quality.

The 2stepQA-based Baseline-3 approach is most relevant as an early Type-001100 DR IQA instantiation. Since 2stepQA combines NIQE and MSSSIM, a direct comparison can be made with the NIQE-MSSSIM DR IQA models by comparing Tables \ref{Table:Baseline_PLCC} and \ref{Table:Baseline_SRCC} with Tables \ref{Table:DR_IQA_Models_PLCC} and \ref{Table:DR_IQA_Models_SRCC}, respectively, where it can be seen that the DR IQA models either perform better than or at par with 2stepQA in a majority of the test cases. However, when other base NR-FR combinations, especially the CORNIA-FSIMc and dipIQ-FSIMc combinations, are adopted, DR IQA Models 1, 2, and 3, outperform 2stepQA nearly comprehensively, and the gaps are large in the most difficult All data cases of Waterloo Exp-II, LIVE MD, and MDIVL databases. These results suggest that the selection of base NR and FR models, and the method of combination (i.e., considering the behavior of multiple simultaneous distortions instead of simple product) are both important in yielding superior performance.

It is also worth noting that the DR IQA Models 1, 2, and 3, are developed by using DR IQA databases V1 and V2, where all training images are annotated by SQB scores~\cite{dnn_database_tip} rather than subjective ratings. Thus, the superior performance of these models on subject-rated test datasets (LIVE MD \cite{db_livemd}, MDIVL \cite{db_mdivl}, and LIVE WCmp \cite{md_2stepQA_jrnl}) has in turn provided a strong demonstration of the value of using SQB \cite{dnn_database_tip} as an alternative IQA data annotation mechanism.

\subsubsection{Inter-Model Comparisons}
\label{DRIQA_PerfInter}

We perform three kinds of inter-model comparisons. First, the complexity and number of parameters increase from the proposed Models 1, 2 to 3, thus presumably, one would expect their performance to improve correspondingly. Somewhat surprisingly, this is not necessarily always the case. Close observation of Tables \ref{Table:DR_IQA_Models_PLCC} and \ref{Table:DR_IQA_Models_SRCC} concludes that in a majority of cases, these models offer similar performance. In particular, Model 1, which is constructed empirically from distortion behavior analysis and uses only 2 parameters, often produces similar performance when compared with the 6-parameter Model 2 and the much more sophisticated SVR-based Model 3. This reveals the value of distortion behavior analysis as discussed in Sections~\ref{sec:MDbehaveAnalysis} and \ref{sec:DRIQAmodeling}. Also note that the performance of Model 1 is not as competitive as the other 2 models in the N-JPG and N-JP2 combinations. This is explained by the construction of Eq.~\ref{mod1-orig} which makes it difficult to adequately capture the complex distortion behavior for these cases as depicted in Figs.~\ref{fig:Plots_ASfd_vs_RSfd} (d) and (e).

Second, we compare across DR IQA architectures. Of the five $\text{AS}_{\text{DR}}$/$\widehat{\text{AS}_{\text{DR}}}$ and $\text{RS}_{\text{FD}}$ combinations, based on the choice of FR/NR methods, the first (FSIMc-FSIMc) belongs to Scenario 1 or Type-100100 DR IQA (Fig. \ref{DRIQA_Scenario1}), while the rest (CORNIA-FSIMc, dipIQ-FSIMc, NIQE-FSIMc, and NIQE-MSSSIM) belong to the more practical Scenario 2 or Type-001100 DR IQA (Fig. \ref{DRIQA_Scenario2}). The main difference is in the availability of the PR image. Since FR methods outperform NR ones and are more reliable, it is natural to expect that the Scenario 1 method should outperform Scenario 2 methods. Interestingly, this is generally not observed in Tables \ref{Table:DR_IQA_Models_PLCC} and \ref{Table:DR_IQA_Models_SRCC}, suggesting that the lack of access to the PR image may not always be critical to DR IQA, as long as appropriate NR and FR methods are used to compute $\widehat{\text{AS}_{\text{DR}}}$ and $\text{RS}_{\text{FD}}$, respectively.

Third, we compare across distortion combinations. Tables \ref{Table:DR_IQA_Models_PLCC} and \ref{Table:DR_IQA_Models_SRCC} report the performance of DR IQA models for seven multiple distortion combinations (Blur-JPEG, Blur-Noise, JPEG-JPEG, Noise-JPEG, Noise-JP2K, NBJ-JPEG, and All data). All models perform quite consistently across the LIVE MD \cite{db_livemd}, MDIVL \cite{db_mdivl}, and LIVE WCmp \cite{md_2stepQA_jrnl} databases, but they do not have all seven combinations. Thus, we focus on the Waterloo Exp-II database \cite{dnn_database_tip}. It appears that the DR IQA models perform quite well for Blur-JPEG, Blur-Noise, and JPEG-JPEG cases, and reasonably well (with few exceptions) for Noise-JPEG, Noise-JP2K, NBJ-JPEG, and All data cases when considered independently, but not at the same level of the first three cases. This is understandable given the variations in distortion behaviors demonstrated in Fig.~\ref{fig:Plots_ASfd_vs_RSfd}. This makes DR IQA an interesting and challenging problem that demands future investigations, especially for the challenging distortion combinations and the All data case.

\section{Conclusion}
\label{sec:Concl}

We make one of the first attempts to establish a DR IQA paradigm, targeting at the problem when only a reference image of degraded quality is available when assessing the quality of a multiply distorted image, a problem that is of practical importance in many real-world applications such as image/video distributions. We lay out possible architectures of DR IQA, introduce a 6-bit code to denote various configurations, and focus on two specific architectures or scenarios. We establish first-of-their-kind large-scale synthetically annotated databases dedicated to DR IQA, and conduct a novel multiple distortion behavior analysis for two-stage distortion pipelines. We also develop novel DR IQA models and make extensive comparisons with different types of baseline models. The results suggest that DR IQA may offer significant performance gain in multiple distortion environments against existing FR and NR IQA paradigms, thereby establishing DR IQA as a novel IQA paradigm in its own right. We hope the current work can inspire significant future research that explores different DR IQA architectures and distortion combinations.

\section*{Acknowledgment}
This work was supported in part by the Natural Sciences and Engineering Research Council (NSERC) of Canada.

\bibliographystyle{IEEEtran}
\bibliography{IEEEabrv,bib_allother,bib_database,bib_DNN,bib_friqa,bib_mdiqa,bib_metricfusion,bib_nriqa,bib_nss,bib_stateval}

\end{document}